\documentclass[preprint, 12pt]{aastex}
\usepackage{epsfig, amsmath, amsfonts}

\newcommand\bb[1] {   \mbox{\boldmath{$#1$}}  }

\newcommand{\dd}[2]{\frac{{\rm d} #1}{{\rm d} #2}}

\def\dd{\partial}

\def\beq{ \begin{equation} }
\def\eeq{ \end{equation} }
\def\spose#1{\hbox to 0pt{#1\hss}}  %from Scott Tremaine
\def\ltsim{\mathrel{\spose{\lower.5ex\hbox{$\mathchar"218$}}
\raise.4ex\hbox{$\mathchar"13C$}}}
\def\gtsim{\mathrel{\spose{\lower.5ex\hbox{$\mathchar"218$}}
\raise.4ex\hbox{$>$}}}

\slugcomment{}

\begin{document}

\title{\bf\LARGE Differential Rotation in Fully Convective Stars}
\author{ Steven A. Balbus\altaffilmark{1,2},
Nigel O. Weiss\altaffilmark{1,3}}

\altaffiltext{1}{Laboratoire de Radioastronomie, \'Ecole Normale
Sup\'erieure, 24 rue Lhomond, 75231 Paris CEDEX 05, France
  \texttt{steven.balbus@lra.ens.fr}}

  \altaffiltext{2}{Adjunct Professor, Department of Astronomy, University of Virginia,
  Charlottesville VA 22903, USA}

  \altaffiltext{3}{DAMTP, Centre for Mathematical Sciences, Wilberforce Road,
   Cambridge CB3 0WA, UK}

\begin{abstract}
Under the assumption of thermal wind balance and effective entropy
mixing in constant rotation surfaces, the isorotational contours
of the solar convective zone may be reproduced with great fidelity.
Even at this early stage of development, this helioseismology 
fit may be used to put a lower bound on the midlatitude {\em radial}
solar entropy gradient, which in good accord with standard 
mixing length theory.  In this paper,
we generalize this solar calculation to fully convective stars
(and potentially planets), retaining the assumptions of
thermal wind balance and effective entropy mixing in isorotational surfaces.
It is found that each isorotation contour
is of the form $R^2 = A+B\Phi(r)$, where $R$ is the radius from
the rotation axis, $\Phi(r)$ is the (assumed spherical)
gravitational potential, and $A$ and $B$ are constant along the contour.
This result is applied to simple models of fully convective stars.
Both solar-like
surface rotation profiles (angular velocity decreasing toward the
poles) as well as ``antisolar'' profiles (angular velocity
increasing toward the poles) are modeled; the latter bear some
suggestive resemblance to numerical simulations.  We also perform
exploratory studies of zonal surface flows similar to those seen in Jupiter
and Saturn.  In addition to providing a practical
framework for understanding the results of large scale numerical simulations, 
our findings may also prove useful
in dynamical calculations for which a simple but viable model for the
background rotation profile in a convecting fluid is needed.  
Finally, our work bears directly
on an important goal of the CoRoT program: to elucidate
the internal structure of rotating, convecting stars.

\end{abstract}


\maketitle
\section{Introduction}

Recent work (Balbus 2009, hereafter B09;
Balbus et al. 2009, hereafter BBLW) suggests that,
away from inner and outer boundary layers, the isorotation
contours in the bulk of the solar convective zone (hereafter SCZ) 
correspond to the characteristic
curves of the vorticity equation in its quasi-linear ``thermal wind'' form.
Since both the entropy $S$ and angular velocity
$\Omega$ figure in this single equation, the mathematical existence of
these characteristics requires that there be some functional 
relationship between $S$ and $\Omega$.

B09 put forth the possibility that this relationship was a consequence
of the SCZ being marginally stable to axisymmetric
magnetobaroclinic modes.  In BBLW, however, it was shown that an
entropy/angular velocity relationship is expected to be present under
very general conditions; a magnetic field is not a prerequisite. 
Because robust convective structures lie in surfaces in which
$\Omega$ is constant, and because these structures mix 
entropy very efficiently,
surfaces of constant entropy and surfaces of angular velocity correspond
with one another.  More precisely, it is not quite the entropy that is
homogenized by the convective mixing, it is the ``residual entropy:''
some radial entropy profile must be present to trigger the
convection, and it is the {\em residual entropy}, the entropy that departs
from the underlying radial profile, that is homogenized by the 
convective mixing in 
constant $\Omega$ surfaces (BBLW).  The confluence of constant 
angular velocity and constant residual entropy surfaces ensures that
there is a functional relationship between them, and the existence
of this functional relationship, even in the absence of knowledge
of its precise form, allows one to deduce the structure of the 
isorotation characteristics
of the thermal wind equation (hereafter TWE).  
If isorotational surfaces are the same as constant residual entropy
surfaces, this is a result of considerable practical interest:
in Appendix II, we show that it may be used in conjunction with helioseismology 
data to place a lower bound on the logarithmic radial
entropy gradient (an otherwise observationally inaccessible quantity)
of a few times $10^{-6}$.  This number is good accord with mixing length
theory estimates (e.g. Stix 2004)\footnote{A recent helioseismic
inversion calculation
by Brun et al. (2009) reports very large entropy variations in the SCZ,
inconsistent with previous numerical simulations, thermal wind balance,
and standard mixing length theory.  These results may, however, be noise
dominated.}.

Because the calculations presented in this paper depend strongly 
on the assumptions of thermal wind balance and the confluence of
constant rotation and residual entropy surfaces, we review the
supportive arguments that apply in environments similar to
the SCZ.  There are four distinct strands:

\noindent{\em Numerical Simulations.}  
The first piece of evidence is provided by numerical
simulations.  A comparison of figure (2e) (isorotational contours) and
(3a) (residual entropy) in Miesch, Brun, \& Toomre (2006; see also figure
[2] of BBLW) shows a similarity that is striking to the eye.  By way
of contrast, there is no such correspondence between surfaces of total
entropy and angular velocity (e.g., figure [2] in BBLW).  The validity
of TWB is evident in figure (5) of Miesch et al.\ (2006).  

\noindent {\em Dynamics.}   The second line of argument is 
based on a very simple physical picture put forth in BBLW, 
and has already been
touched upon above.  Localized, embedded disturbances in a shearing
system will be distorted into surfaces of constant $\Omega$.  This
follows from mass conservation alone, and is most easily shown by
using coordinates comoving with the background flow (BBLW).  
But the embedded disturbances that are
of interest here are convective
rolls, which mix (and thus homogenize) {\em residual} entropy.
Since the convective rolls are at once in surfaces of constant
rotation and constant residual entropy, these two types of surfaces
must correspond.

\noindent{\em Agreement between theory and observations.}
Even if a functional relationship between the angular velocity and
residual entropy were simply postulated {\em ad hoc,} the resulting
agreement between the calculated isorotational contours and the
helioseismology data is so striking that the fit itself is an independent piece
of evidence for the reasoning on which it is based (cf. fig.\ [1]
of BBLW).  
The relationship traansforms the TWE from an equation relating
certain partial derivatives of $\Omega$
and $S$ into a self-contained, quasi-linear
partial differential equation for $\Omega$ alone.  
The isorotation contours then emerge 
as the characteristics of this equation.  
These characteristics thus depend for their very
existence on a functional relationship between (residual) entropy and
angular velocity.   
Without a functional relationship between entropy
and angular velocity, the notion of isorotation characteristics would
make no mathematical sense.  With the functional relationship in
place, the agreement between analytic theory and observations is remarkable.

\noindent{\em Heuristics.}   A fourth line of argument depends only upon the
the mathematical structure of the TWE (in its undeveloped form relating
$S$ and $\Omega$) and the 
obeservation that the gradient of the SCZ angular velocity is
dominated at mid-latitudes by its latitudinal angular component.
We defer the details until the following section,
when the TWE is explicitly presented.  

Any of the four arguments on its own would motivate a study of
convection and differential rotatation in which surfaces of 
constant angular velocity and constant residual entropy coincided.
Taken as a whole, they comprise a truly compelling picture.

%Once posited, this functional relationship ultimately
%leads to both an exceptionally simple analytic expression for the isorotation
%characteristics.  To compare this result with the data, 
%at a given location in the convective zone, each characteristic is
%uniquely determined by matching the slope of the helioseismolgy $\Omega$
%contour at the chosen location.  The resulting fit, figure (1) in BBLW,
%speaks for itself.  
%

We are thus motivated to extend the mathematical technique of B09 to 
fully convective
stars, under the assumption that solar-like dynamical conditions
prevail.   In practice, this means solving the TWE not just for the $1/r$ 
gravitational potential appropriate to the SCZ 
($r$ is the usual spherical radius),
but for arbitrary potentials $\Phi(r)$.   We present a suite of 
solutions
for the resulting isorotation contours under a variety of simple
assumptions for the mathematical form of the entropy/angular momentum
relationship as well as for the type of boundary condition used for 
calculating $\Omega$.  
We consider interior angular velocity
solutions for cases in which
the surface rotation is monotonically increasing from pole to equator,
monotonically decreasing in the same direction, and
for zonal surface flows (seen, for example,
in Jupiter and Saturn) as well.  We also treat problems in which
the angular velocity is given along the axis of rotation, in which case
there may be very little surface variation.

The profiles computed should provide a framework within which one
may better understand numerical simulations
of rotation in convecting stars and planets.  Moreover,
if the rotation profiles of at least some fully convective stars are not 
dominated by magnetic fields, or turbulent or meridional
fluxes, the calculated isorotation contours 
are viable predictions for the gross interior features of these
bodies.  Extracting such information from
the global p-mode data of fully convective stars
is a primary goal of the CoRoT satellite mission.  

An outline of our paper is as follows.  Section 2 is
a presentation of the mathematical solution for the isorotation
contours of a fully convective star in which thermal wind
balance is a good approximation.  In section 3, we construct
some specific solutions, corresponding to both solar-like 
and ``antisolar'' surface profiles,
followed by models with banded zonal flows.  Section 4 is a 
concluding discussion and summary.

\section{Analysis}

Throughout this paper, we follow the notation of BBLW.  
We denote cylindrical coordinates by radius $R$, azimuthal angle
$\phi$ and axial coordinate $z$.  Spherical coordinates 
are given by $(r, \theta, \phi)$, where $r$ is the radius
from the origin, $\theta$ is the colatitude angle, and $\phi$ is
once again the azimuth.   Unless otherwise stated, $\Omega$, the 
pressure $P$ and the density $\rho$ 
are understood to be azimuthal averages,
independent of $\phi$.  
%The velocity $\bb{v}$ will in general contain 
%convective motions---any averaging will be explicitly discussed.
The dimensionless entropy function $\sigma$ is defined by:
\beq
\sigma \equiv \ln P\rho^{-\gamma},
\eeq
where $\gamma$  is the usual (specific heat ratio) adiabatic index.  
As noted in the Introduction, $\Phi(r)$ will refer to the local gravitational
potential.  For the problem of interest, 
$\Phi(r)$ is a well-known, tabulated function.  

In thermal wind balance, one ignores
contributions from convective turbulence (i.e., the
convective Rossby number is small [Miesch \& Toomre 2009])
and magnetic fields.  This is 
an important simplification that appears to work well for 
the bulk of the convective zone of the Sun      
(between roughly 0.75 and 0.95 solar radii), 
but must of course be reexamined critically for each application.      

The equation of thermal wind balance is (e.g. Thompson et al. 2003,
Miesch 2005, B09):
\beq\label{eq1}
R{\dd\Omega^2\over \dd z}=\left(1\over\gamma r\right) {d\Phi\over dr}{\dd \sigma
\over \dd\theta}.
\eeq
For the SCZ, $d\Phi/dr=GM_\odot/r^2$, where $G$ is the gravitational
constant and $M_\odot$ represents a solar mass.  As in BBLW, we assume that
there is a functional relationship of the form
\beq\label{eq1bis}
\sigma' \equiv \sigma -\sigma_r =f(\Omega^2),
\eeq
where $\sigma_r$ is any function of $r$ (in practice something very close to an
angle-averaged $\sigma$) and $f$ is an unspecified function.   We have
noted earlier that this
mathematical condition will have a physical basis if
convection mixes the residual entropy $\sigma'$ in constant $\Omega$
surfaces.  But the mathematical structure of equation (\ref{eq1})
together with the most striking feature of the solar data already suggest
a very simple heuristic argument (first mentioned in the Introduction)
leading to a relation of the form seen in
equation (\ref{eq1bis}), an argument that is independent of whatever
dynamical mechanism might be responsible for establishing the relationship.
Helioseismology data show that at solar midlatitudes, the
$\theta$ gradient of $\Omega$ dominates over its $r$ gradient.  Now in the
TWE, $\sigma$ is operated upon only by the $\theta$ gradient\footnote{
Likewise, $\Omega$ appears only within a $z$ gradient operator,
and may have a function of $R$ added to it without changing the equation.
But this freedom is lost once $\Omega$ is specified on a particular
surface.}.  Thus, if $\sigma$
happens to have a dominant $r$ gradient (not an unreasonable guess for
the Sun), this component can largely be eliminated by subtracting off a
suitably chosen radial function $\sigma_r$, i.e., by forming a $\sigma'$
quantity.  Substituting $\sigma'$ for $\sigma$ does not change the TWE
at all.  But by its construction, this procedure will 
enhance the importance of the $\theta$ gradient
of $\sigma'$ relative to its $r$ gradient, and therefore it motivates a search
for solutions under the assumption that $\sigma'$ and $\Omega$
share isosurfaces in common.  This is precisely the content of equation 
(\ref{eq1bis}).  

It is of interest to consider why there can be such substantial departures
from the Taylor-Proudman constraint of $\Omega=\Omega(R)$ in a system that is
so nearly adiabatic (and thus barotropic).   The condition that isorotational
and constant $\sigma'$ surfaces coincide provides an immediate answer: if 
$\Omega=\Omega(R)$ then $\sigma'=\sigma'(R)$ and $\dd\sigma'/\dd\theta$
must therefore be present.   If this partial derivative is of order
$(R\Omega/v_s)^2$ or larger, where $v_s$ is the adiabatic sound speed,
it is enough to account for the order unity
departure from Taylor-Proudman columns seen in the helioseismology data and
our calculations.  

We assume that there is a stable, long lived average rotation profile
$\Omega(r,\theta)$ that is present in the star.  The evolutionary
processes that are responsible for the establishment of $\Omega$ are
likely to be complex, involving an as yet poorly understood combination
of mechanical and thermal transport.  For the problem at hand,
however, these
details are less important than the existence of $\Omega(r,\theta)$
itself, and the concomitant assumptions that the profile is in an
average sense time steady and in thermal wind balance.  If either
of these assumptions fails for the fully convective stars we
consider here, our model is no longer valid.

Using equation (\ref{eq1bis}), the TWE (\ref{eq1}) becomes
\beq\label{eq2}
{\dd\Omega^2\over \dd r} - 
\left[
\left(f'\over\gamma r^2\ \sin\theta \cos\theta \right) {d\Phi\over dr} +
{\tan\theta\over r}
\right]
{\dd\Omega^2\over \dd\theta} = 0 ,
\eeq
where $f'=df/d\Omega^2$.
The solution to equation (\ref{eq2}) is that $\Omega^2$ is constant on surfaces
described by
\beq
{d\theta\over dr } = - \left(f'\over\gamma r^2\ \sin\theta \cos\theta \right) {d\Phi\over dr} -
{\tan\theta\over r}.
\eeq
This may be written
\beq
{d\ \over dr} \left( r^2\sin^2\theta +{2f'\over\gamma} \Phi(r) \right)=0, 
\eeq
which has the solution
\beq\label{result}
 r^2\sin^2\theta = A - {2f'\over\gamma} \Phi(r),
 \eeq
where $A$ is a constant of integration.  This is a rather remarkable result,
and worth stating more informally:
{\em if the TWE is obeyed and 
entropy is mixed as efficiently as possible in constant rotation surfaces,
the isorotation contours are of the form $R^2 = A+B\Phi(r)$, where $A$
and $B$ are constants along each contour.  }

If our mathematical problem is to be well-posed,
boundary conditions for $\Omega^2$ must be 
specified.  One simple possibility is
to demand that the angular velocity be
a prescribed function of the colatitude angle
$\theta_0$ on some spherical surface $r=r_0$.  
For the sake of simplicity, we consider
this to be the surface of the star.  
Then the isorotation contour will be given by
\beq\label{iso}
 r^2\sin^2\theta = r_0^2\sin^2\theta_0 - {2f'\over\gamma} [\Phi(r)-\Phi(r_0)].
 \eeq

In some stars, however,
the characteristics may be quasi-spherical, in which case
most of the isorotation contours will not penetrate the surface.  (This seems
to be true, for example, of convective stars with a poleward increasing
surface rotation rate.)
We require a formulation for the characteristics
that is suitable for these cases.  For quasi-spherical isorotation contours,
$\Omega^2$ is most conveniently prescribed as a function of $z_0$
along the axis of rotation.  Then, the contours are defined by the
equation
\beq\label{quasi}
r^2\sin^2\theta =  {2f'\over\gamma} [\Phi(z_0)-\Phi(r)].
 \eeq

Since we will be concerned here with fully convective stars, 
the potential $\Phi(r)$ is that of an $n=1.5$ polytrope, where $n$ 
(the polytropic index)
is given by $1/(\gamma-1)$.  
To solve for $\Phi(r)$, recall that the equation of
hydrostatic equilibrium may be immediately integrated for a 
polytropic star, giving (e.g. Schwarzschild 1958):
\beq\label{eq3}
{\gamma\over\gamma-1}{P\over\rho} +\Phi(r)=-{GM\over r_0},
\eeq
where $M$ is the mass of the star, and 
the integration constant on the right side is chosen so
that the stellar potential matches the external potential $-GM/r$
at the stellar surface $r=r_0$.  
For the polytropic equation of state
$P=K\rho^\gamma$, where $K$ is a constant defined by the adiabat,
equation (\ref{eq3}) becomes
\beq\label{eq4}
(n+1) K\rho^{1/n} +\Phi(r) = -{GM\over r_0}.
\eeq

To complete our determination of $\Phi(r)$, we need an explicit solution
for $\rho^{1/n}$.   This is given by (Chandrasekhar 1939)
\beq
\rho^{1/n} = \rho_c^{1/n} \Theta_n ({\xi}),
\eeq
where $\rho_c$ is the central density of the star, $\xi$ is the dimensionless radius
\beq\label{lea}
\xi = {r\over a}, \quad a^2 = {K(n+1)\rho_c^{(1/n)-1}\over 4\pi G}
\eeq
and $\Theta_n(\xi)$ is the
Lane-Emden function of order $n$, which satisfies
\beq
{d^2(\xi \Theta_n)\over d\xi^2} = - \xi (\Theta_n)^n.
\eeq
Extensive tables of $\Theta_n$ are available in the literature (e.g. Horedt 2004);
for our needs a simple Pad\'e approximant compiled by Pascual (1977) will suffice (see Appendix I).

Returning to the problem at hand, solving equation (\ref{eq4}) for $\Phi$ now gives
\beq
\Phi(r) = -{GM\over r_0} - (5/2) K\rho_c^{1/n} \Theta_{3/2}(\xi).
\eeq
The zero of $\Theta_{3/2}(\xi)$ is located at $\xi_0 = 3.65375$
(Chandrasekhar 1939), corresponding to the surface $r=r_0$.  
Our final expression for the isorotation contour (\ref{iso}) is given by
\beq\label{eqc}
r^2\sin^2\theta = r_0^2\sin^2\theta_0 + {3f' c_S^2 }\Theta_{3/2}(\xi),
\eeq
where $c_S^2$ ($=K\rho_c^{1/n}$)
is the square of the isothermal sound speed at the center of the star, and 
we have used $\gamma=5/3$.  The coefficient of $\Theta_{3/2}$ will be of order
$r_0^2$ if the dimensionless quantity $\Omega^2f'\sim
(r_0\Omega/c_S)^2 \ll 1$.  
The corresponding isorotation contour associated with equation (\ref{quasi}) is
\beq\label{quasi2}
r^2\sin^2\theta =  {3f' c_S^2 }[\Theta_{3/2}(\xi)- \Theta_{3/2}(z_0\xi_0/ r_0)],
\eeq
where $z_0$ refers to the value of $z$ at the start of the characteristic
along the axis of rotation.  The ratio $z_0/r_0$ thus runs between $0$ and $1$.

For solar-like surface profiles, 
$f'$ will generally be negative.  This is because 
convection is more efficient along the rotation axis, so the
poles will be associated with higher temperatures and small
rotation rates relative to the equator.  For antisolar
profiles, similar reasoning suggests that $f'$ should be positive.
It is not yet
possible to say more than this without detailed numerical simulations
of the turbulent flow in convective stars.  In the
spirit of B09 and BBLW, we therefore introduce the function
$F(\sin^2\theta_0)$, which is of orer unity, rewriting equation (\ref{eqc})
as
\beq
(r/r_0)^2 \sin^2\theta =(\xi/\xi_0)^2\sin^2\theta = 
\sin^2\theta_0 + F(\sin^2\theta_0)\Theta_{3/2}(\xi),
\eeq
where
\beq
 F(\sin^2\theta_0)\equiv {3f' c_S^2 \over r_0^2}<0
 \eeq
is some simple dimensionless function of $\sin^2\theta_0$.  
In practice, we will limit ourselves to linear 
or constant functions, there being at this stage no particular
exigency for greater mathematical sophistication.

For solar-like surface profiles,
let us assume that $F(\sin^2\theta_0)=-\beta_1-\beta_2\sin^2\theta_0$.  
Then, solving for $\sin^2\theta_0$ leads to 
\beq\label{sin2}
\sin^2\theta_0 = { (\xi/\xi_0)^2\sin^2\theta +\beta_1 \Theta_{3/2}(\xi)\over
1 - \beta_2 \Theta_{3/2}(\xi)}.
\eeq
If $\Omega^2$ is now a specified function at the 
surface, $\Omega_0^2(\sin^2\theta_0)$ say,
the solution throughout the stellar interior is given 
by substituting for $\sin^2\theta_0$
from equation (\ref{sin2}) in the argument of $\Omega_0^2$.  
This is the desired solution when $\Omega_0$ is specified initially on a spherical
surface.

For the antisolar surface profiles,
the $z_0$ inversion
associated with equation (\ref{quasi2}) is somewhat complicated, except when $F$ is
a (positive) constant, and we will restrict ourselves in this work to this case.
We then find
\beq\label{z02}
\left( z_0\xi_0\over r_0\right)^2 = \Lambda(X), \quad X\equiv\Theta_{3/2}(\xi) -
{\xi^2\over \xi_0^2 F}\sin^2\theta,
\eeq
where the (very simple) $\Lambda$ function, essentially an inversion of $\Theta_{3/2}$,
is presented in Appendix I.  Given the rotation profile along $z_0$, $\Omega_0^2(z_0^2)$
say, the solution throughout the stellar interior is given by substituting for
$z_0^2$ from equation (\ref{z02}) in the argument of $\Omega_0^2$.

\section {Contour description}
\subsection{Solar surface profiles}

We choose the surface rotation profile $\Omega(\sin^2\theta_0)$
to be equal to be that of the Sun,
a simple Taylor expansion in $\sin^2\theta_0$:
\beq\label{omsol}
(\Omega_0/2\pi) = (386.2 + 198.7\sin^2\theta_0 - 66.7\sin^4\theta_0).        
\eeq
The units on the right are nano-Hz.  
Representative solutions are shown in figures 1--4.

\begin{figure}
\epsfig{file=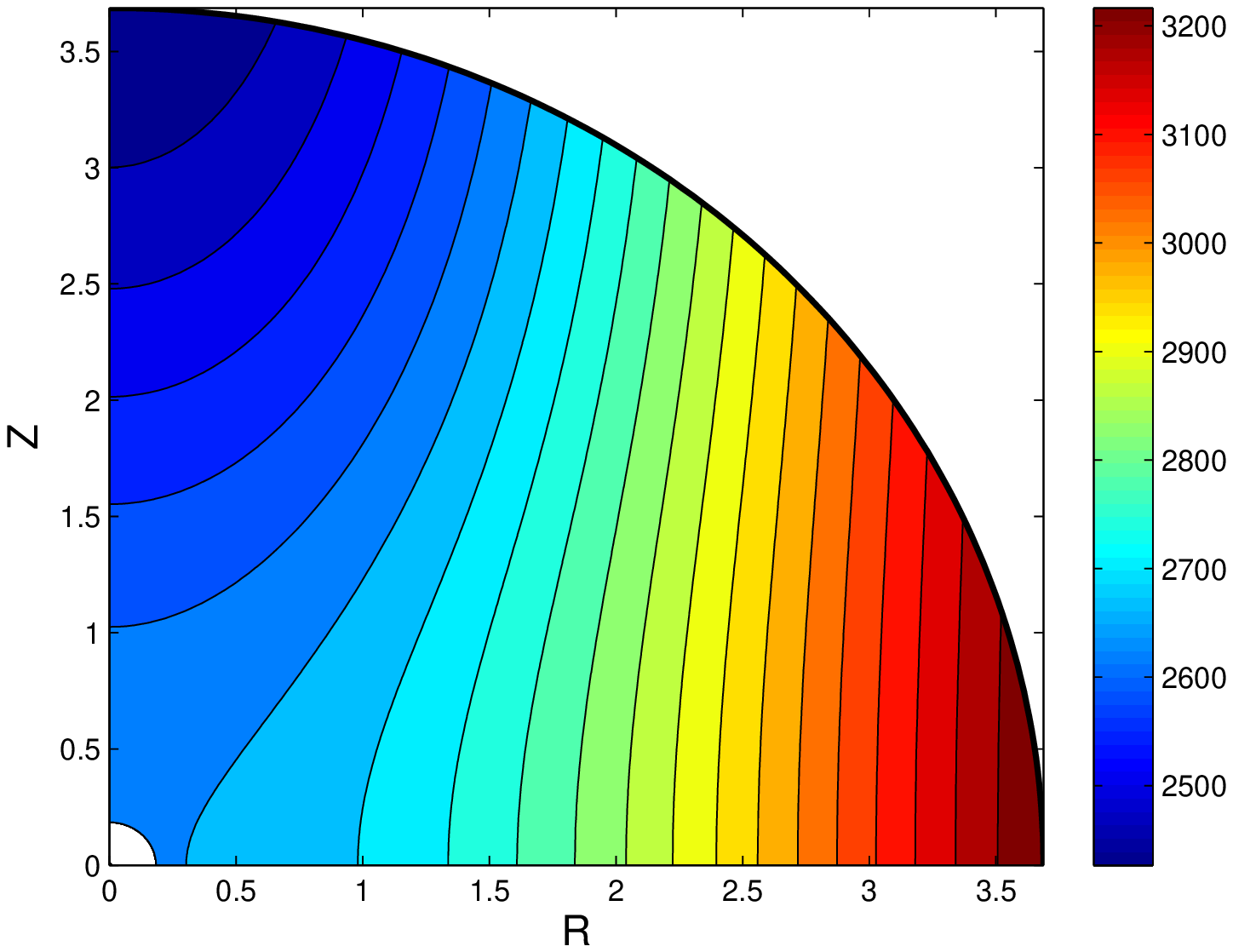, width=8 cm}%,angle=-90}
\epsfig{file=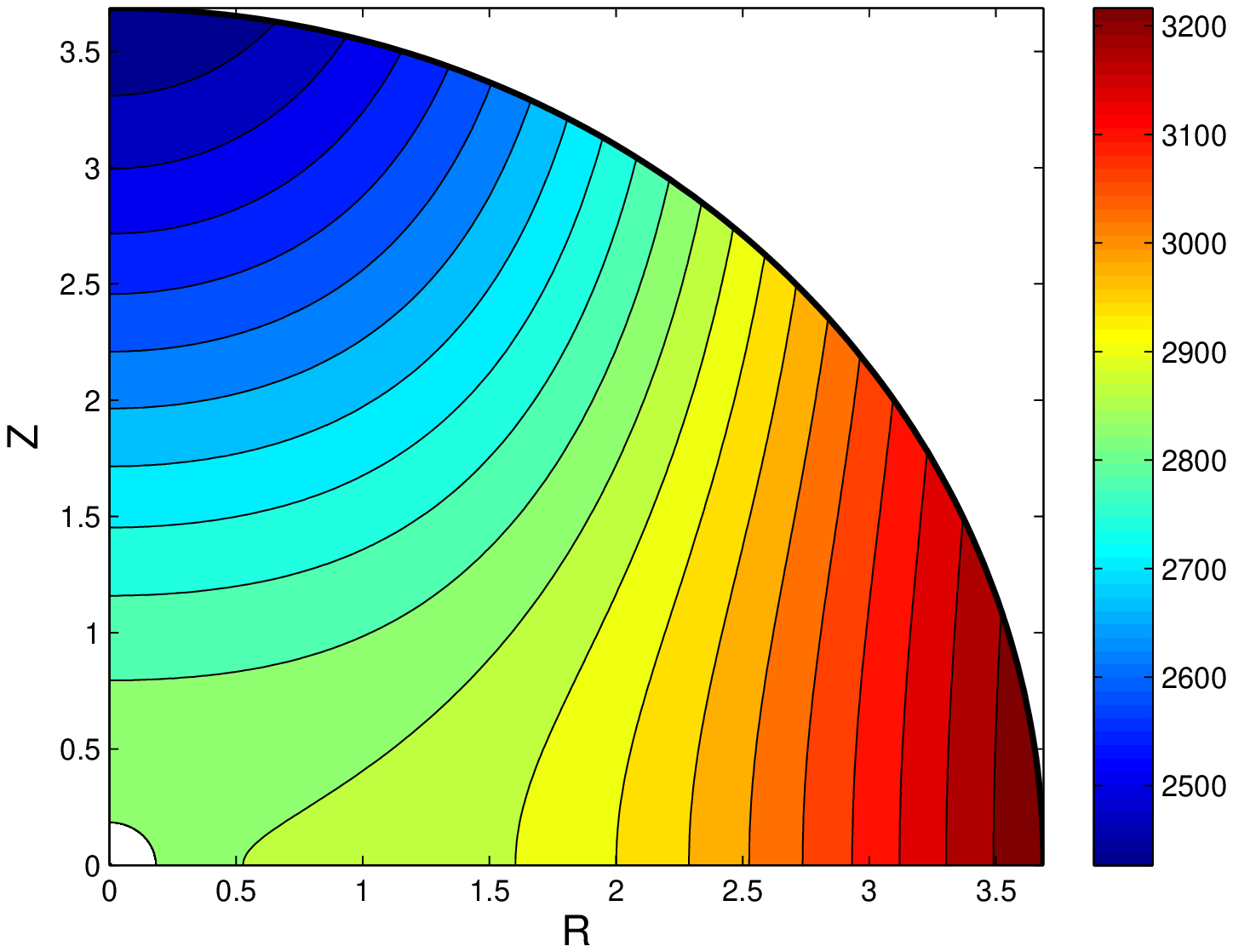, width=8 cm}%,angle=-90}
\caption{Isorotation contours for a solar surface
rotation profile.  In all figures, the abscissa $R$ and 
ordinate $Z$ are given
in units of the Lane-Emden radius $a$ (eq. [\ref{lea}]).
$\beta_1=0.2$, $\beta_2=0$ (left); 
$\beta_1=0.4$, $\beta_2=0.0$ (right).} 
\end{figure}
\begin{figure}
\epsfig{file=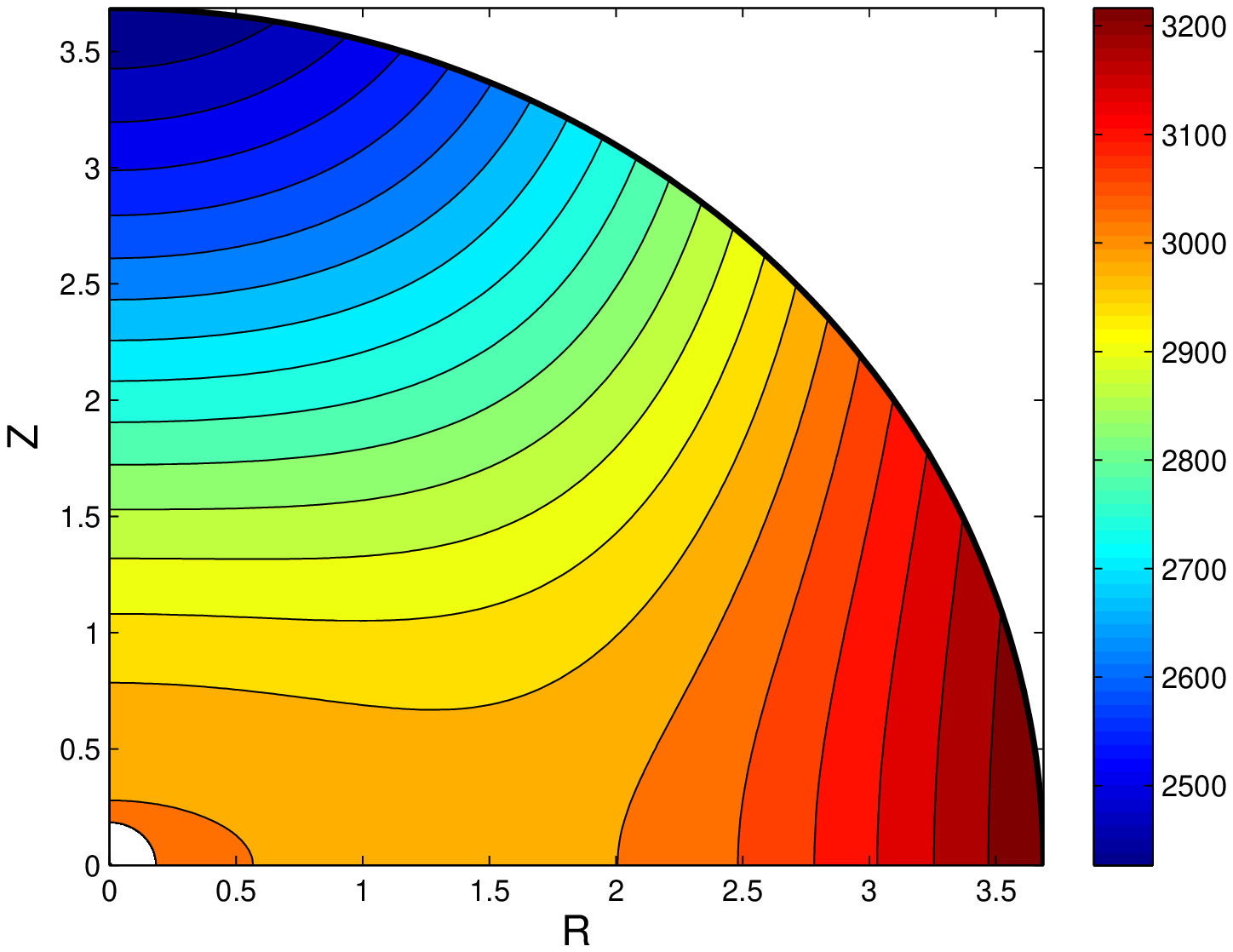, width=8 cm}%,angle=-90}
\epsfig{file=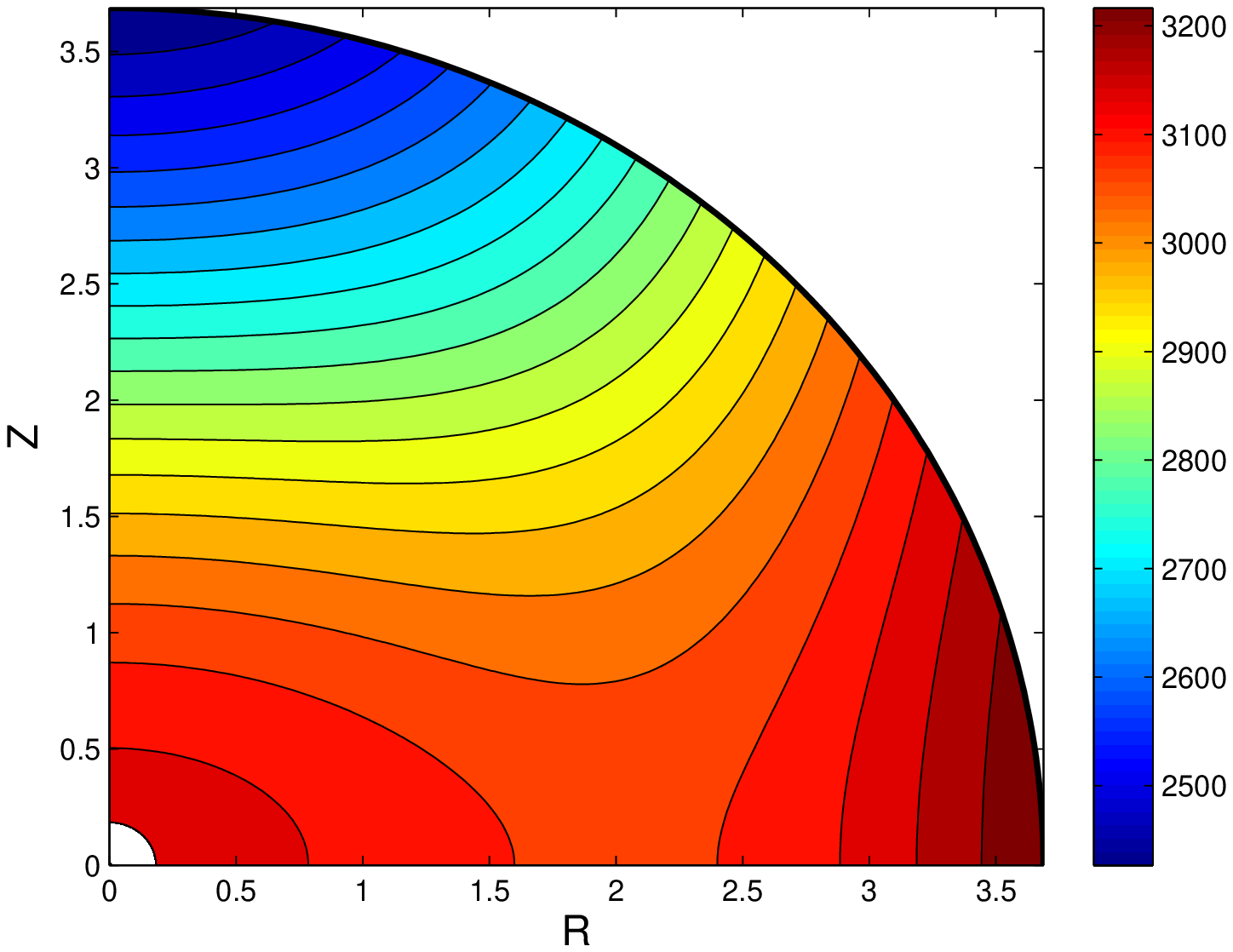, width=8 cm}%,angle=-90}
\caption{As in figure(1), with $\beta_1=0.6$, $\beta_2=0.0$ (left); 
$\beta_1=0.8$, $\beta_2=0.0$ (right).} 
\end{figure}

\begin{figure}
\epsfig{file=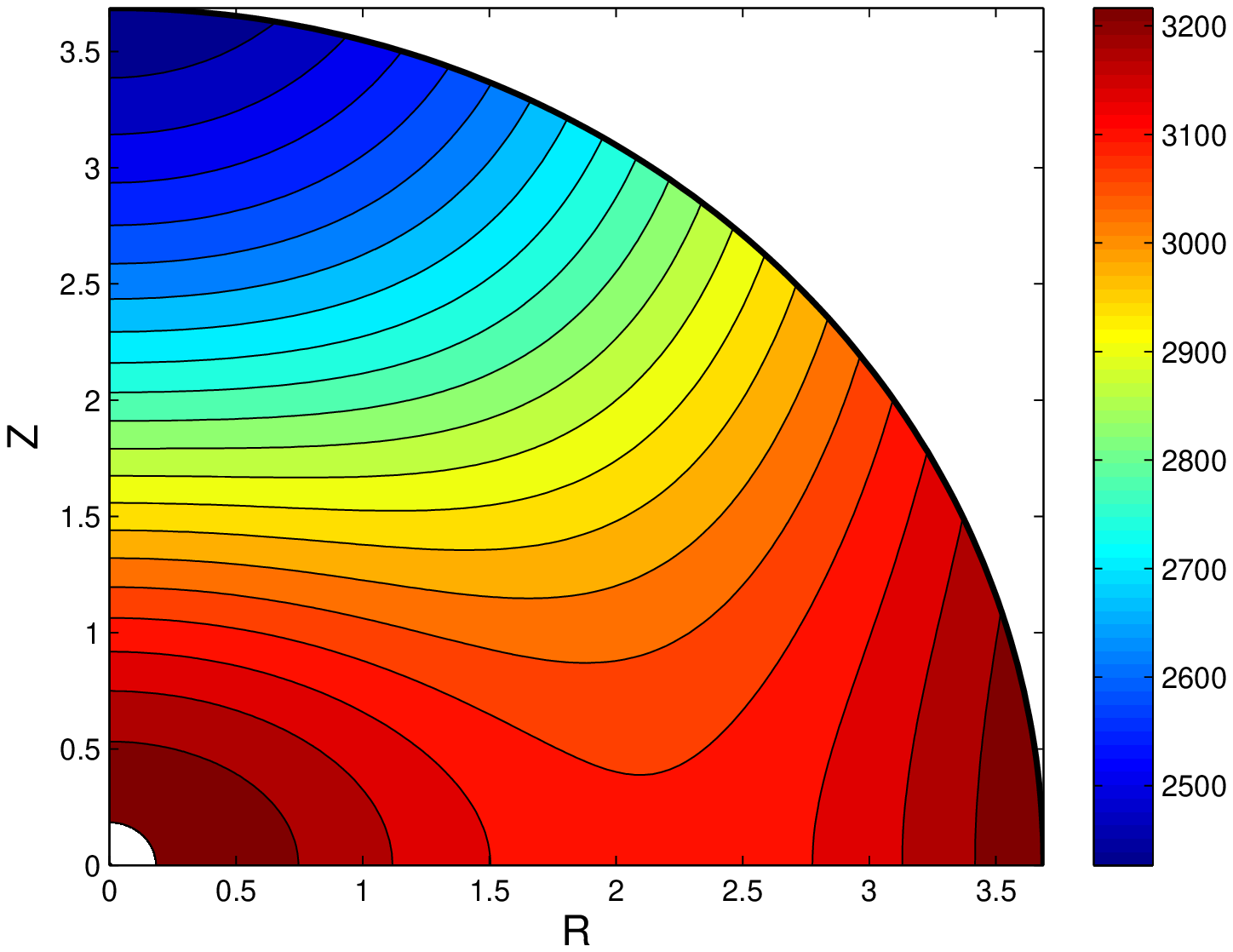, width=8 cm}%,angle=-90}
\epsfig{file=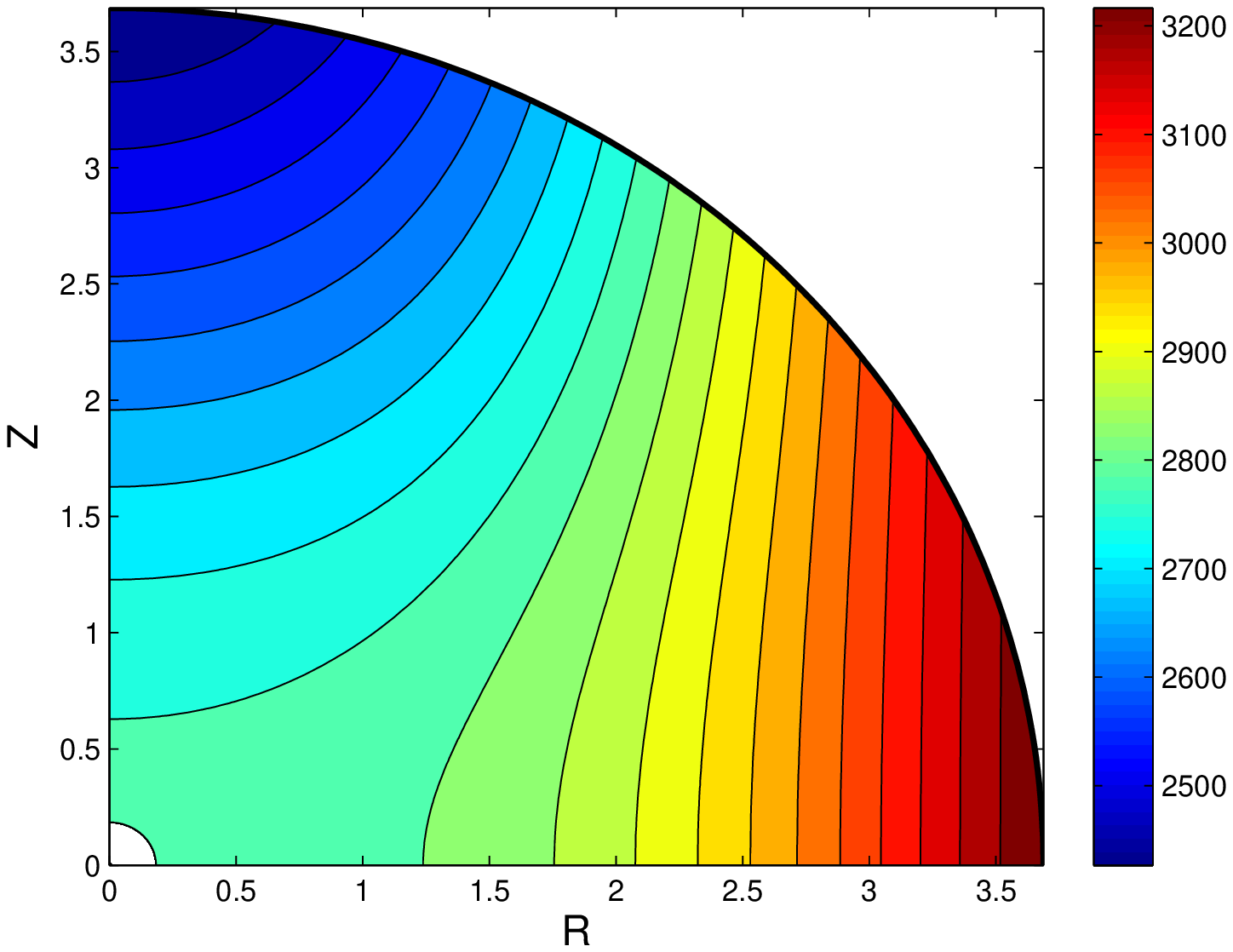, width=8 cm}%,angle=-90}
\caption{As in figure (1), with $\beta_1=0.5$, $\beta_2=0.5$ (left); 
$\beta_1=0.5$, $\beta_2=-0.5$ (right).} 
%\end{figure}
%\begin{figure}
\break
\epsfig{file=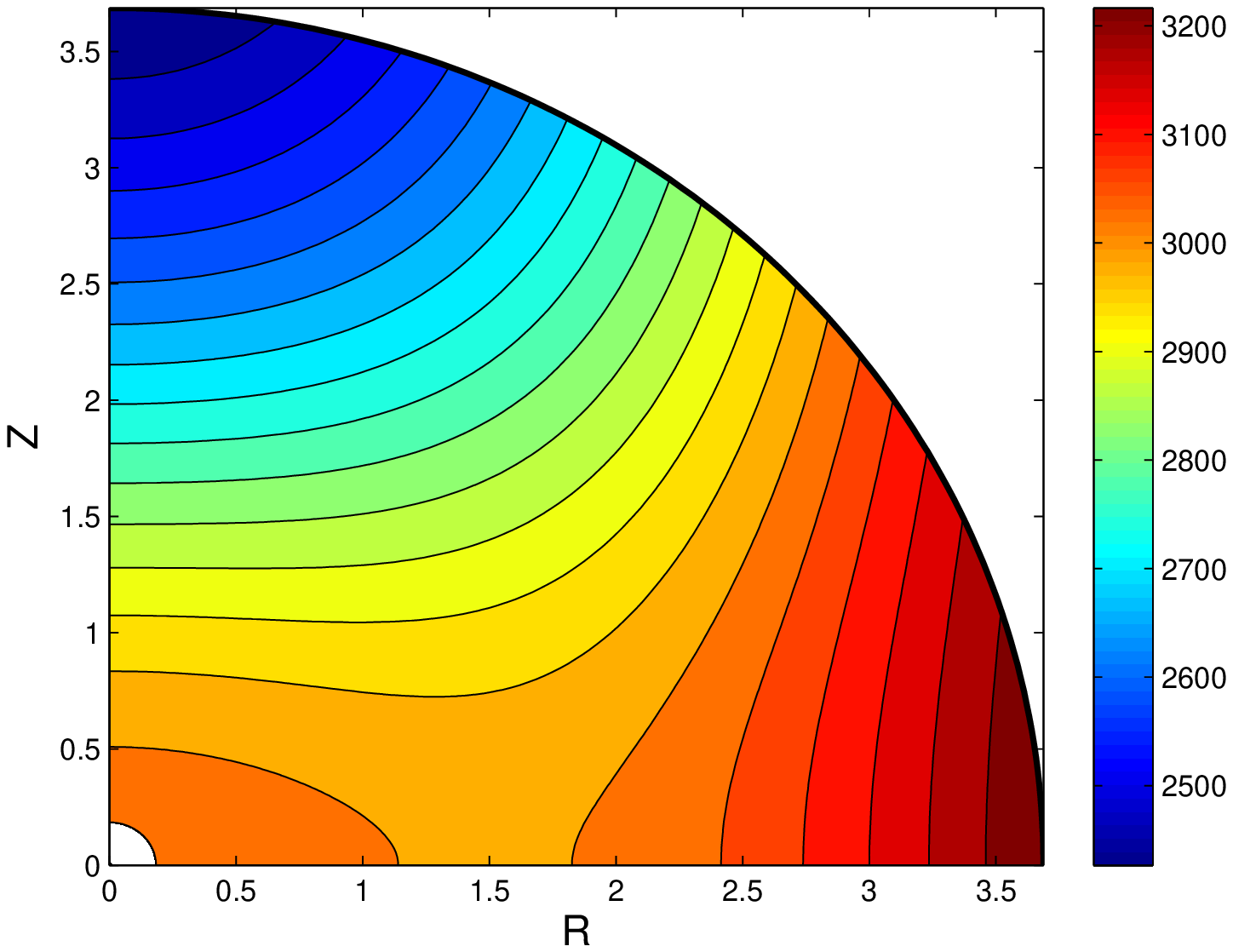, width=8 cm}%,angle=-90}
\epsfig{file=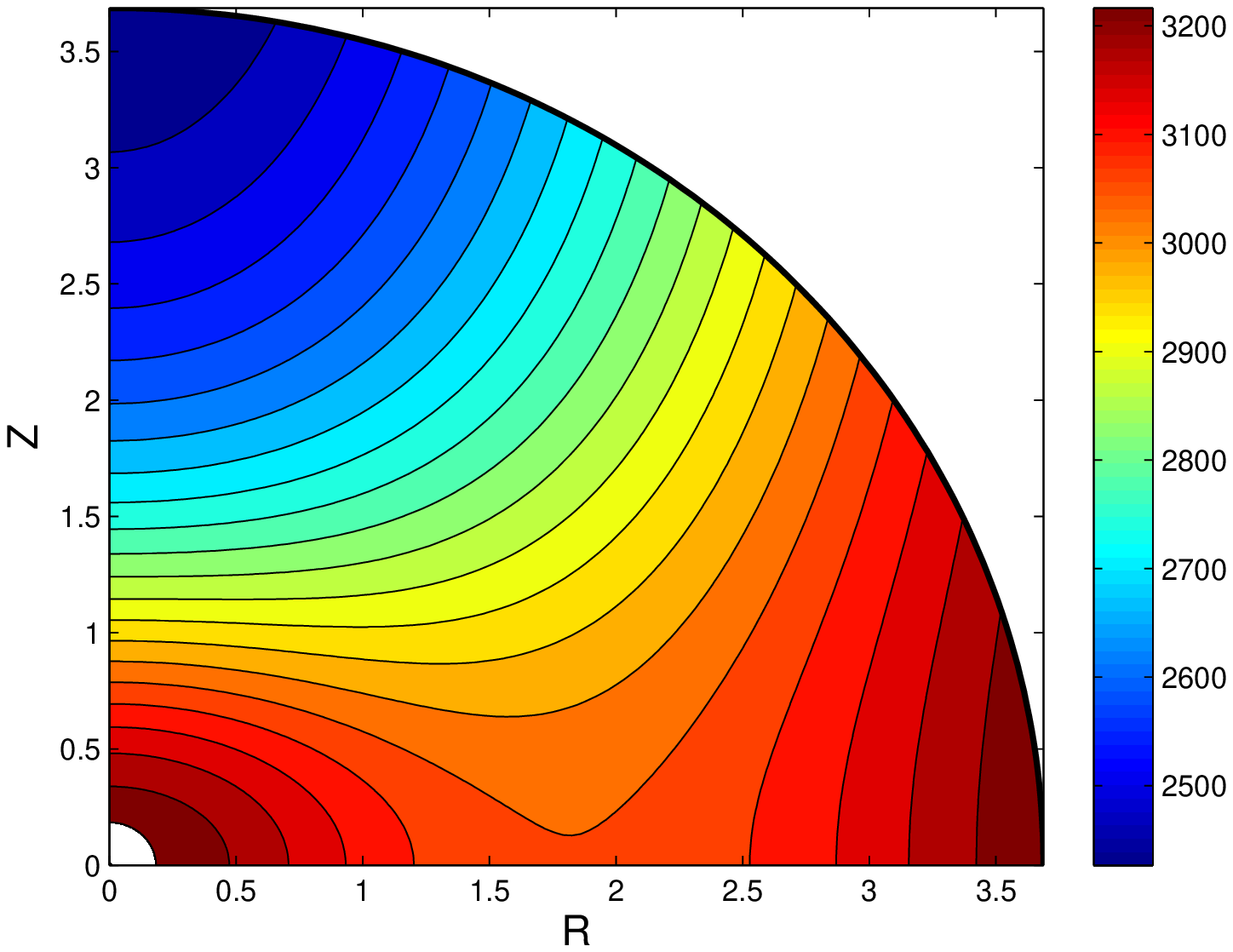, width=8 cm}%,angle=-90}
\caption{As in figure (1), with $\beta_1=0.5$, $\beta_2=0.2$ (left); 
$\beta_1=0.2$, $\beta_2=0.8$ (right).} 
\end{figure}

The four profiles of figures 1 and 2 correspond to a suite of 
increasing values of $\beta_1$, $(0.2, 0.4, 0.6, 0.8)$
that show the contours evolving from nearly cylindrical
Taylor-Proudman columns (corresponding to the limit
$\beta_1=\beta_2=0$),
to a much more ``flat-bottomed'' configuration.
In figures 3 and 4, we allow
$\beta_1$ and $\beta_2$ to vary separately, as indicated in the
captions.  Aside from a tightening or loosening of the contours,
there is not a great qualitative difference between the cases of
vanishing versus finite $\beta_2$.  

The isorotation contours are, roughly speaking, divided into two classes,
which we will term {\em polar} and {\em equatorial}.   Polar contours extend
from the surface, and cross through the rotation axis. 
Equatorial contours, by contrast, extend from the surface and 
cross through the equatorial plane.   In some cases, there are contours
that remain within the star, never reaching the surface.  These tend to be
quasi-spherical, deep in the interior
near the stellar core.  Mathematically, this corresponds
to values of $\sin^2\theta_0$ in excess of unity, but the contours
themselves are not unphysical: they simply remain below the surface of
the star, and are valid solutions of the TWE.  

There is a particular value for $\theta_0$, say $\theta_c$, which 
is the critical divide.  Contours with $\theta_0<\theta_c$ are polar,
while contours with $\theta_0>\theta_c$ are equatorial.  In no
cases do contours cross: caustics are absent from the solutions
that we have studied.  This is {\em not} true for the $1/r$ potential
used in the calculations of B09 and BBLW.  In that case,
caustics formed when the
solutions were extended beneath the convection zone.  The Lane-Emden
function $\Phi_{3/2}(\xi)$, on the other hand,
is evidently sufficiently well-behaved at small
$r$ to avoid the formation of caustics in the isorotation contours.  

Perhaps the most striking feature of these plots
is how reminiscent of the Sun they are, despite the different
gravitational potential and the extended domain.  The contours show the
familiar basic pattern of lying in planes of constant $z$
near the axis of rotation, becoming much more cylindrical near the equator,
and quasi-radial in the transition region between.  Closer to
the stellar surface, we see the characteristic solar feature of 
parallel lines once again in evidence.  We have of course
chosen a solar-like surface profile, but the results are generic for any
slowly varying, poleward decreasing, surface profile with 
reflection symmetry about the equator.

\subsection{Antisolar surface profiles.}

We next relax the constraint of a poleward decreasing surface rotation
profile.  Poleward increasing profiles are seen in simulations of
convective red giant stars (Brun \& Palacios 2009).   Since
thermal convection is expected still to be most efficient parallel
to the rotation axis (where it is unencumbered by the
Coriolis force), the function $f'(\Omega^2)$ is now 
{\em negative.}   This causes interesting changes in the 
geometry of the isorotation contours.

\begin{figure}
\epsfig{file=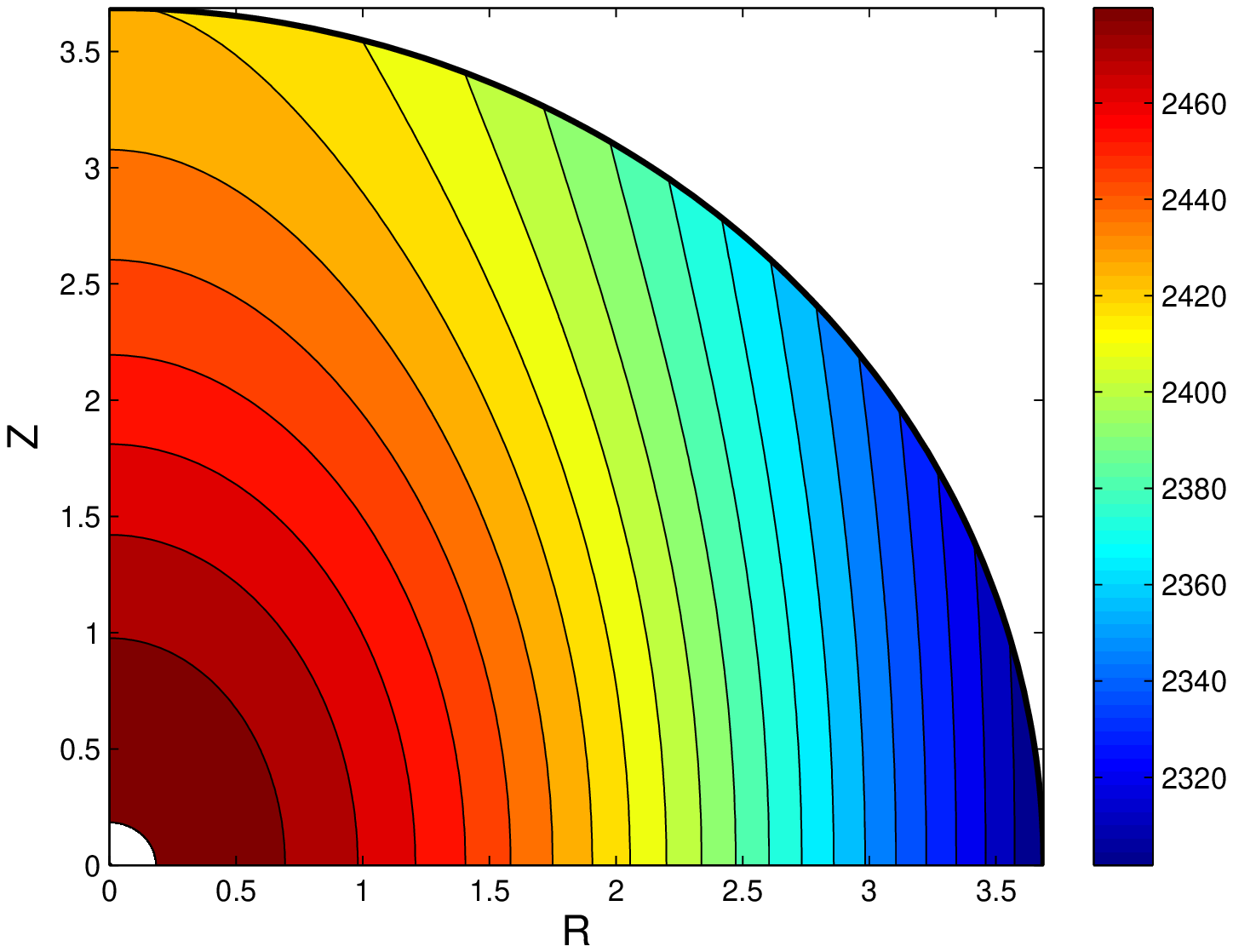, width=8 cm}%,angle=-90}
\epsfig{file=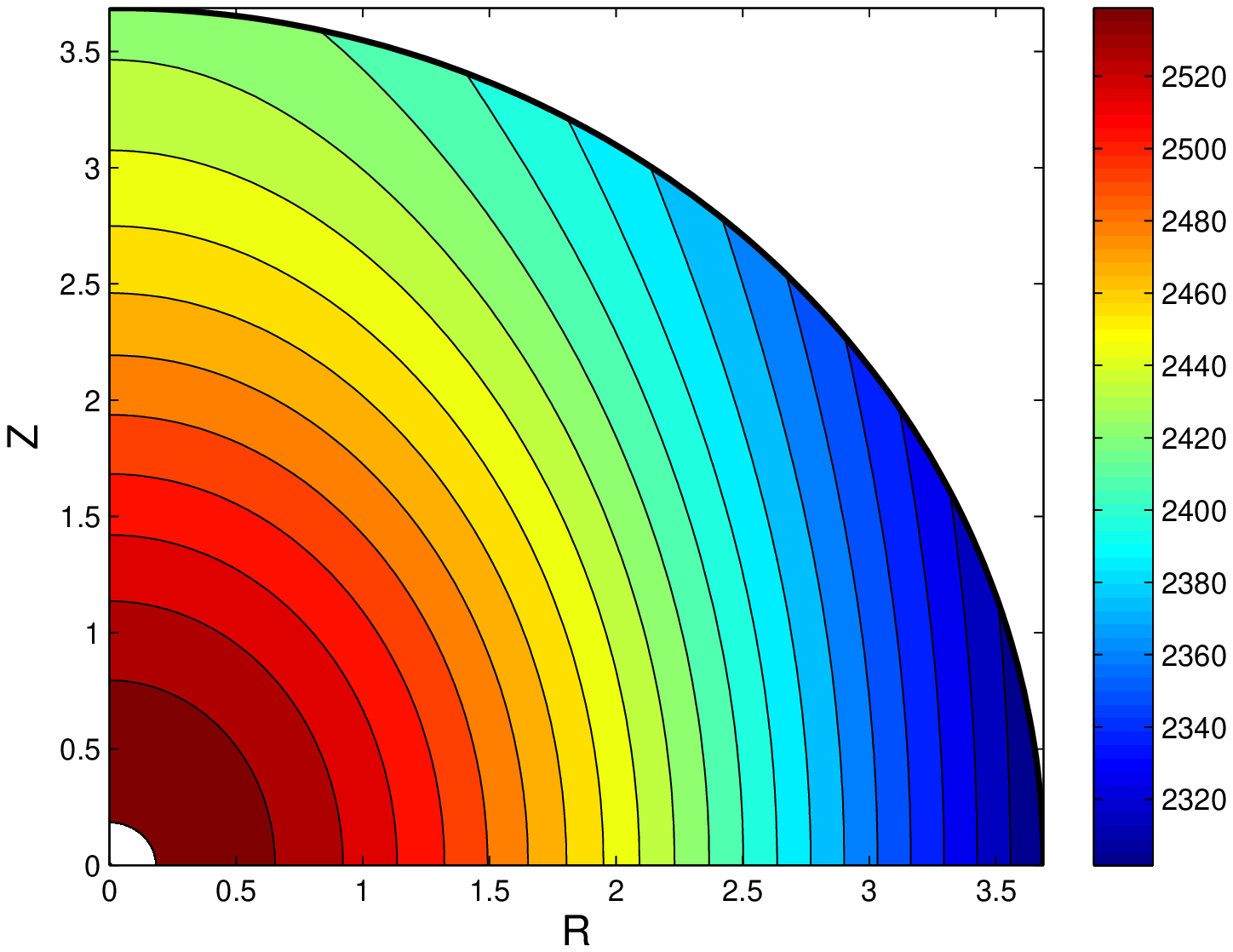, width=8 cm}%,angle=-90}
\caption{ Isorotation contours for an antisolar surface rotation
profile (eq. [\ref{omantisol}]). As in figure (1), the abscissa $R$ and
ordinate $Z$ are given
in units of the Lane-Emden radius $a$ (eq. [\ref{lea}]).
$\beta_1=-0.5$, $\beta_2=0$ (left);
$\beta_1=-1.0$, $\beta_2=0.0$ (right).}
\end{figure}
\begin{figure}
\epsfig{file=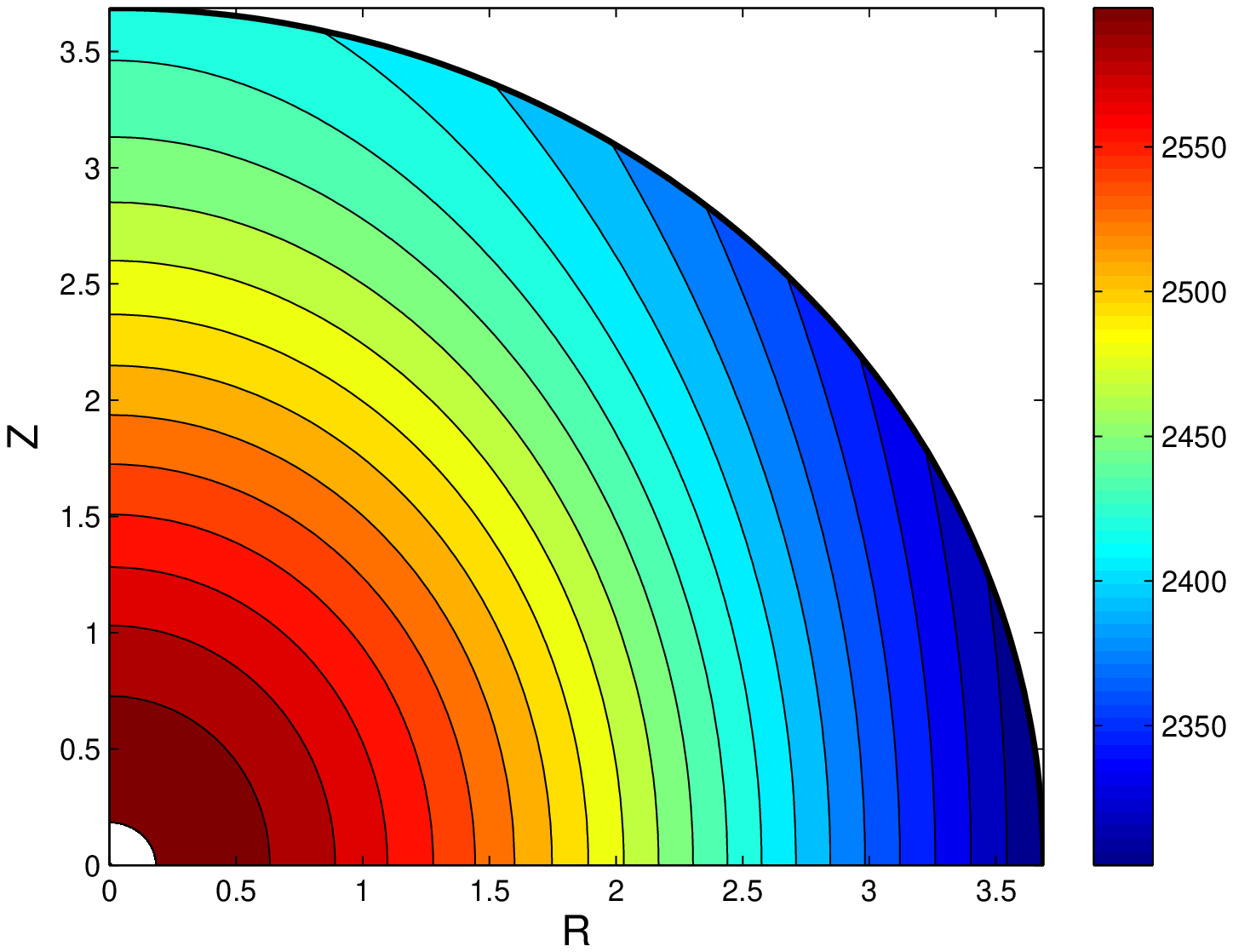, width=8 cm}%,angle=-90}
\epsfig{file=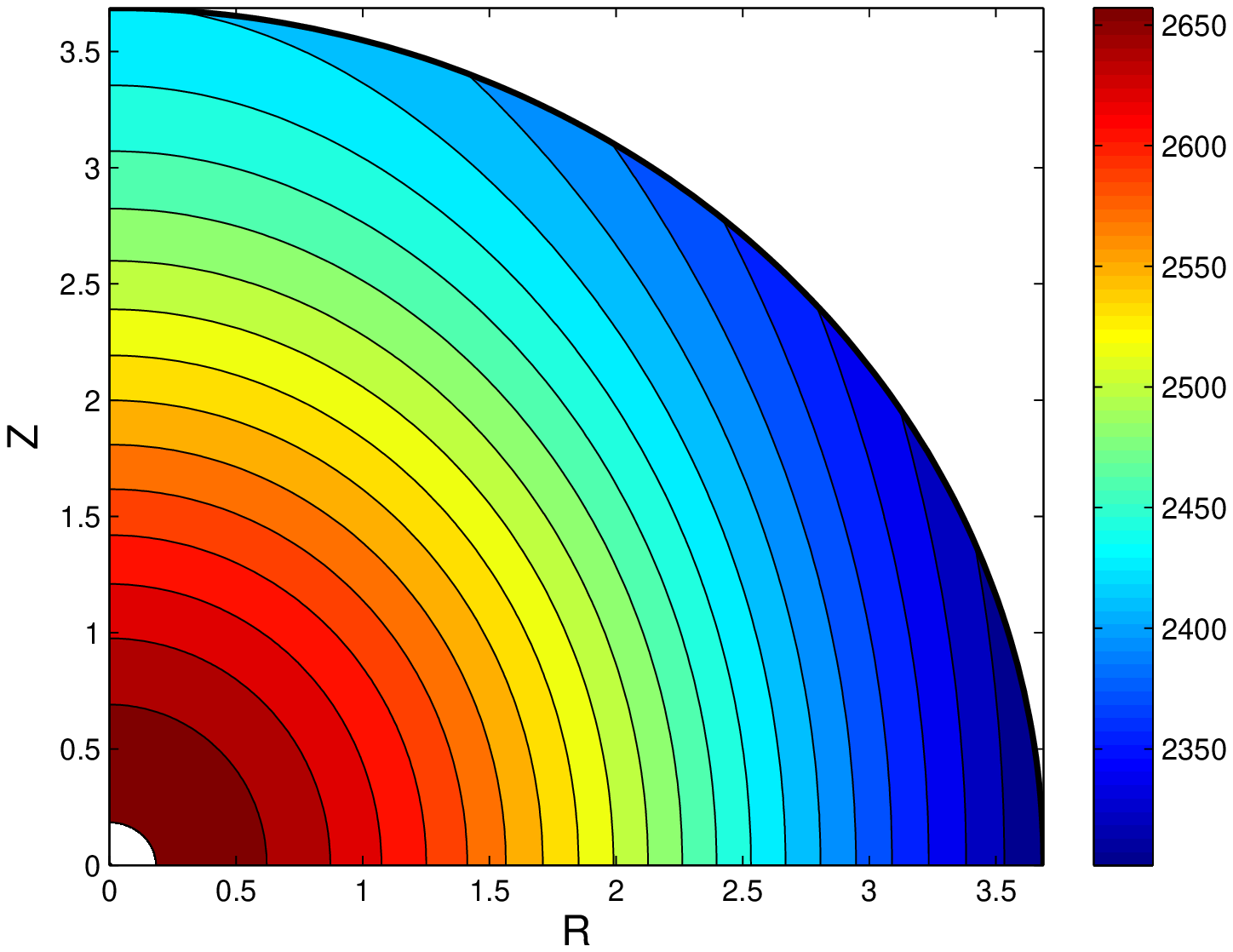, width=8 cm}%,angle=-90}
\caption{As in figure (5), with $\beta_1=-1.5$, $\beta_2=0.0$ (left);
$\beta_1=-2.0$, $\beta_2=0.0$ (right).}
\end{figure}

\subsubsection{$\Omega$ initially specified on the surface}

Figures (5) and (6) show a suite of internal isorotation
contour diagrams for a poleward
increasing surface profile:
\beq\label{omantisol}
(\Omega/2\pi) = 386.2 - 19.87\sin^2\theta_0.
\eeq
Compared with equation 
(\ref{omsol}) for 
our fiducial solar poleward
decreasing profile, we have changed the sign of the
$\sin^2\theta_0$ term and reduced its magnitude by a factor of ten.
We display four values of $\beta_1$,
$-0.5, -1, -1.5, -2$, with $\beta_2$ set equal to $zero$, so
that $f'$ is a global constant.  The isorotation contours
in this case do not resemble the Sun at all, for they now
have a different topology.   There is, for example, no
critical contour dividing isorotation curves that start
at the surface and then head either for the axis or for the 
equator.  Instead, all contours cross the equatorial plane,
and curve in the same sense (toward the rotation axis) as they 
rise upward from the equator.
The most strongly inwardly curving contours first find the rotation axis
(where they end), the
less strongly curved contours encounter the surface before the axis. 
This is reminscent of the slowly rotating red giants in the Brun \& Palacios 
(2009) simulations.   In the numerical studies, however, the surface is 
in near uniform rotation and is best approached using the formalism
of equation (\ref{quasi}) (see below).  
For small values of $|\beta_1|$, the contours
look, not surprisingly, cylindrical.  This is clearly the Taylor-Proudman
regime.  But as the magnitude of $\beta_1$ increases, the contours
become more spherical (prolate).  In contrast to the case of solar surface
profiles (and the Sun itself), in which the rotation rate increases
with distance from the axis, the antisolar surface profiles
are associated with an interior rotation structure with $d\Omega^2/dR <0$.

\begin{figure}
\epsfig{file=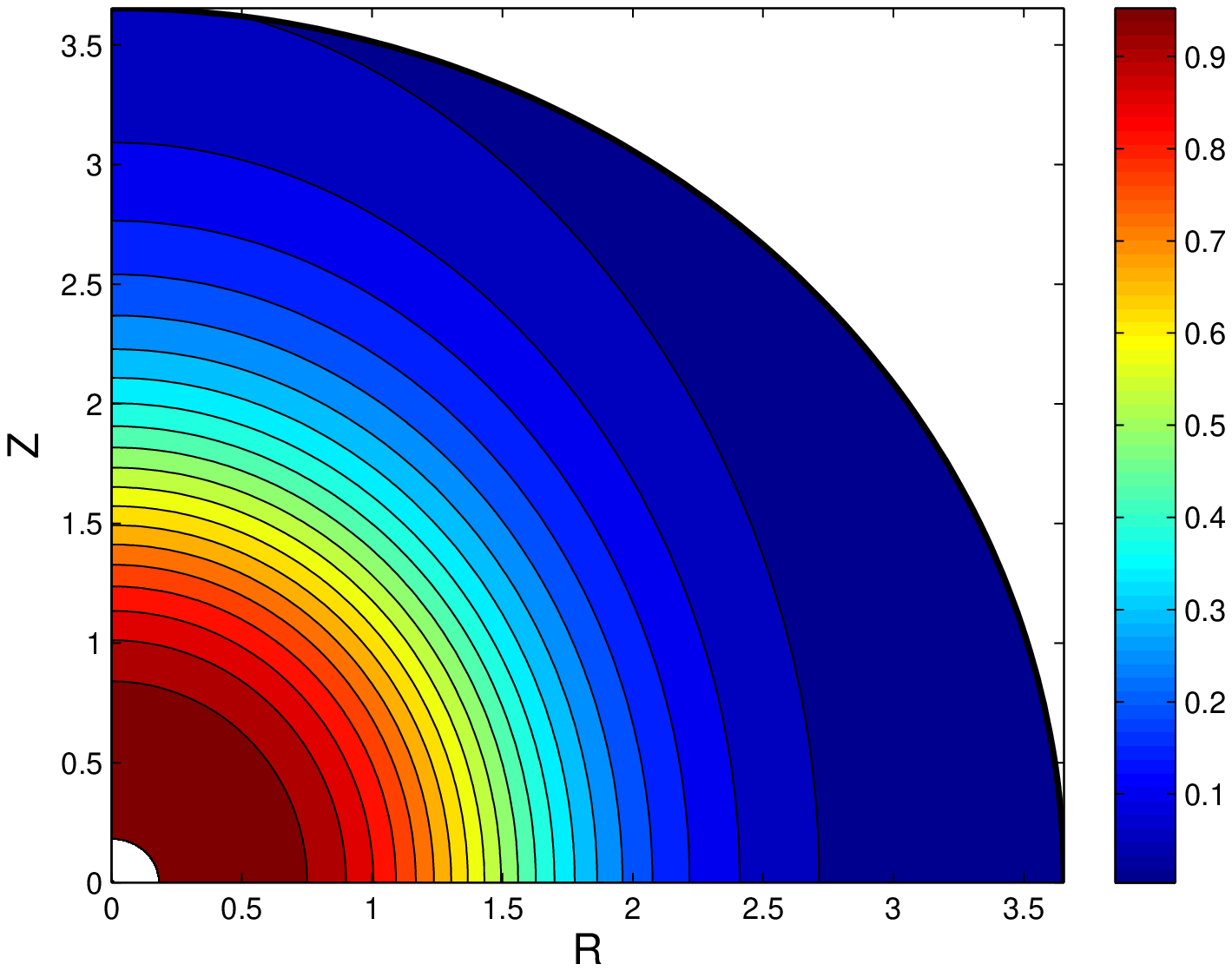, width=8 cm}%,angle=-90}
\epsfig{file=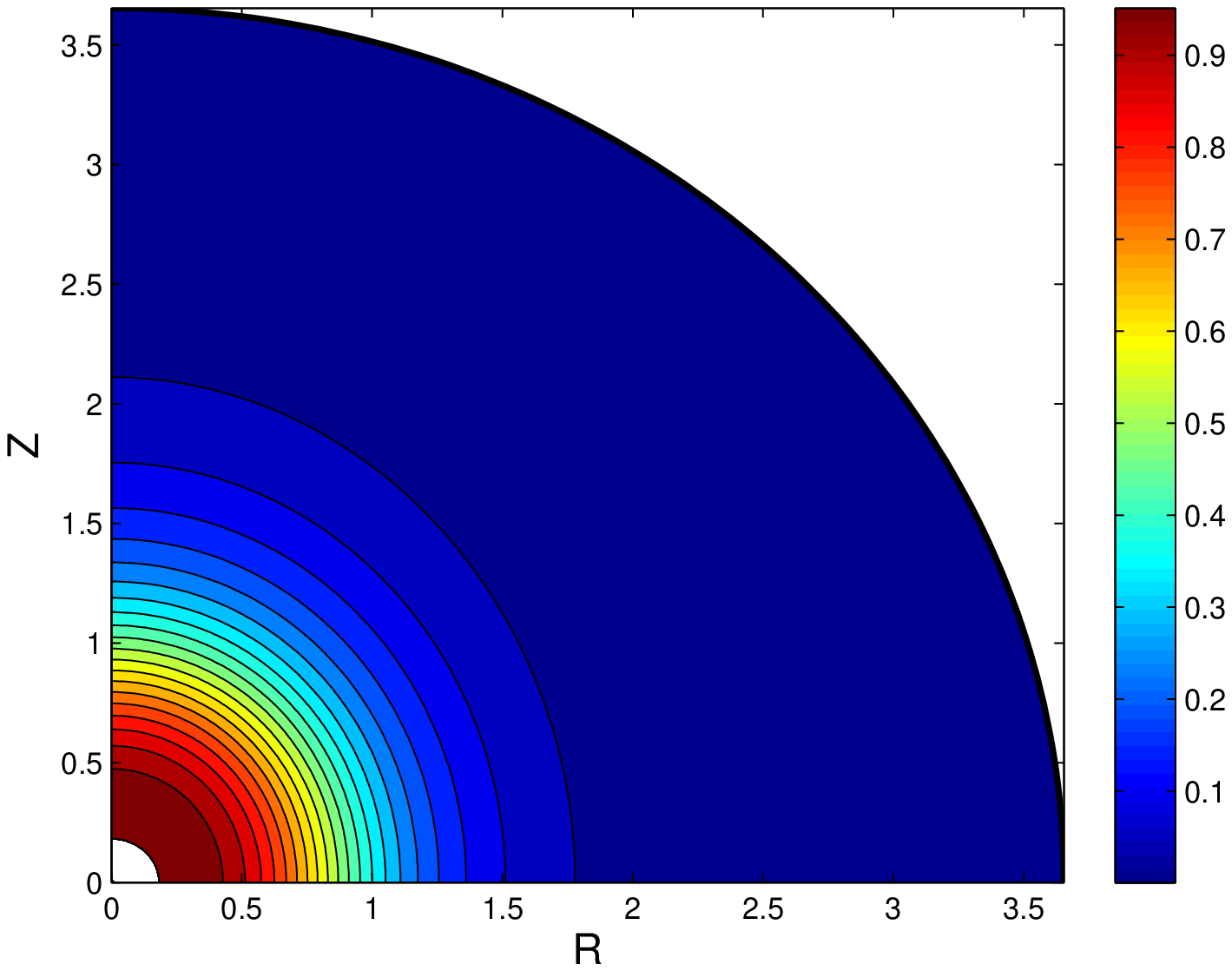, width=8 cm}%,angle=-90}
\caption{Internal isorotation contours for a convective
star with $\Omega$ specified along the rotation axis. The
solution is generated by equations (\ref{quasi2}), 
(\ref{z02}), and (\ref{axis}).  The abscissa $R$ and
ordinate $Z$ are given
in units of the Lane-Emden radius $a$ (eq. [\ref{lea}]).
$F=2$ and $\beta=0.1$
(left) and $1$ (right).}
\end{figure}
\begin{figure}
\epsfig{file=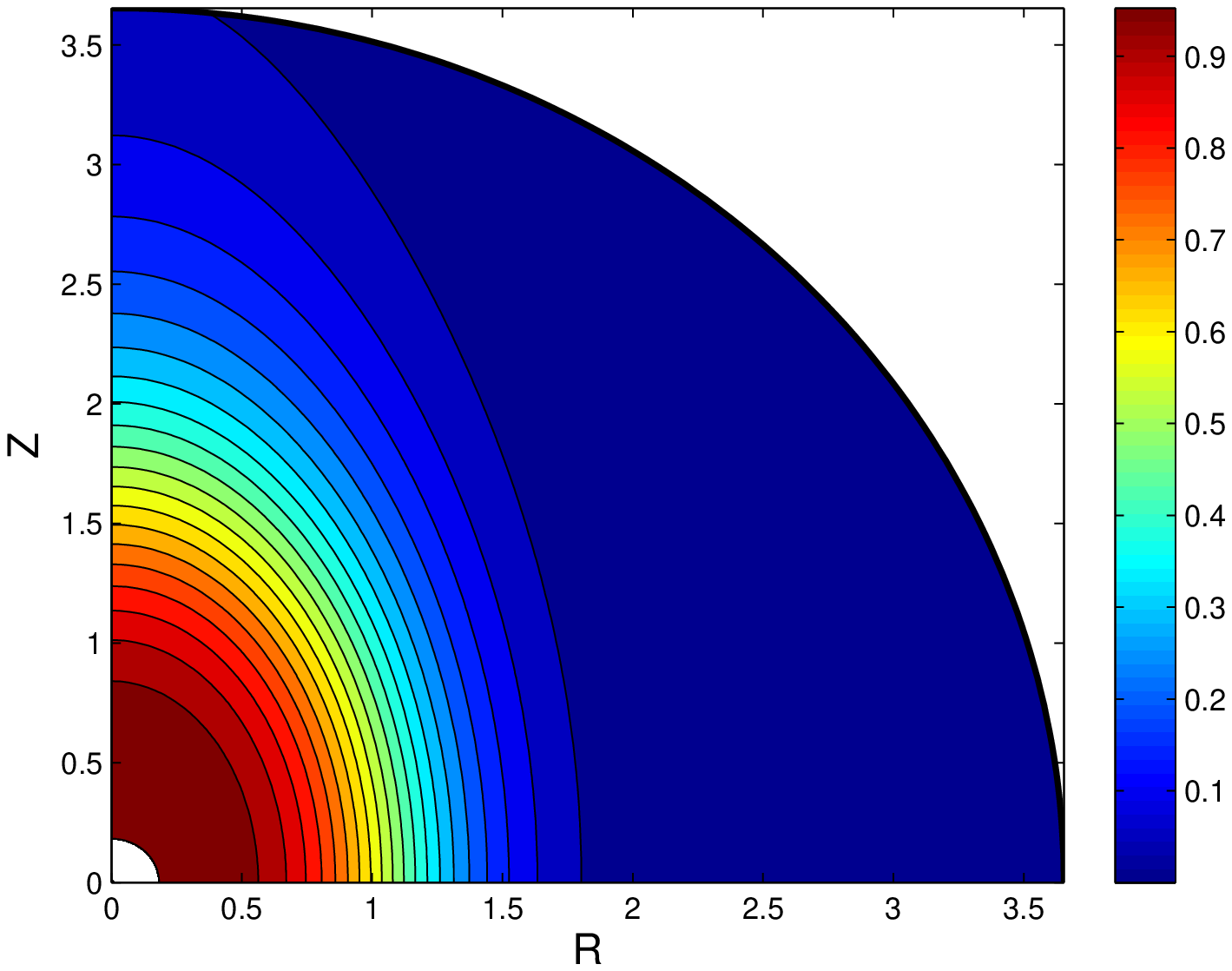, width=8 cm}%,angle=-90}
\epsfig{file=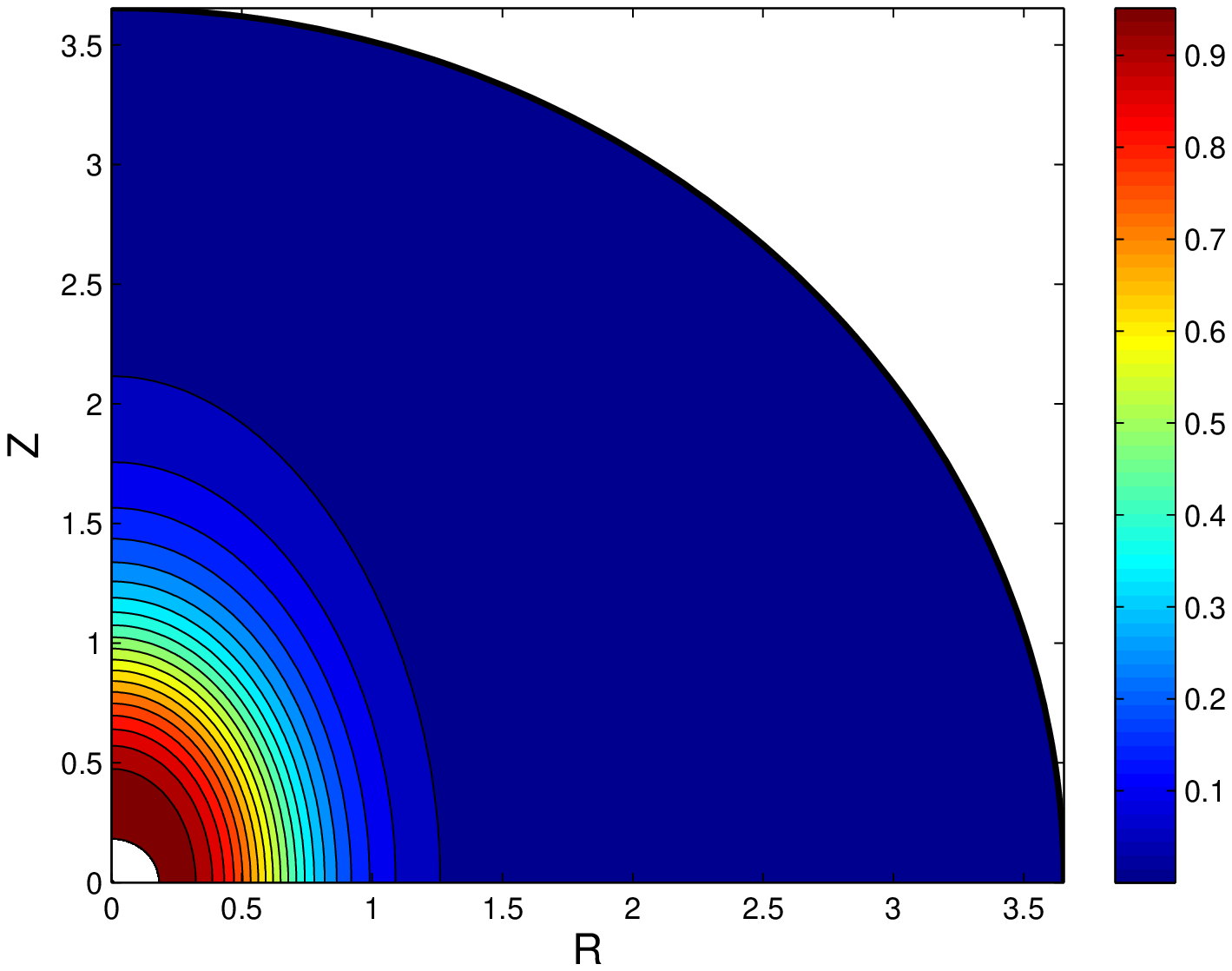, width=8 cm}%,angle=-90}
\caption{As in figure (7), with $F=0.4$, $\beta=0.1$
(left), $\beta=1$ (right).}
\end{figure}

\subsubsection{$\Omega$ initially specified along the rotation axis}

In stars showing significantly more $\Omega$ variation along the axis of
rotation than along a surface meridional contour, it makes sense to
consider the initial rotation function to be of the
form $\Omega_0(z_0^2)$, and use the formulation embodied in equations
(\ref{quasi2}) and (\ref{z02}).   We have adopted the model
\beq\label{axis}
\Omega_0(z_0^2) = \left[1+\beta z_0^4\right]^{-1},
\eeq
leaving $\beta$ as a parameter.  Specifying the function $F$
(treated here as a constant) and $\beta$ then uniquely
defines the model.  

Figures (7) and (8) show a selection of results.  In figure (7),
we have set $F=2$.  On the left, $\beta=0.1$; on the right,
$\beta=1$.  In figure (8), $F=0.4$ with $\beta=0.1$ on the left
and $\beta=1$ on the right.  The larger value of $F$ in figure (7)
corresponds to a {\em smaller} rotation rate, and the isorotation
contours are more rounded.  Figure (8) corresponds to a larger rotation
rate, and the contours are more elongated along the axis, distorted
in the direction of Taylor-Proudman columns.   Depending on the value
of $\beta$, the rotation can either be confined to a central core or
extend to the surface.  These findings compare very favorably with the 
morphologies found in the numerical simulations of Brun \& Palacios
(2009).  

\subsection{Zonal surface flows}

Going further afield, one may speculate on what 
our theory would predict for the internal rotation profile of a convecting
body whose surface exhibited zonal flow---as is seen in the giant
planets of our solar system.  This truly is
a speculative domain, since even the sign of the $d\sigma'/ d\Omega^2$
is not certain for this class of flows.  Our motivation is simply to note 
an alternative to Taylor-Proudman
columns for the internal flow contours of
giant convecting gaseous planets and possibly brown dwarfs.
In the case of Jupiter and Saturn the surface flows are
likely to be shallow (e.g., Liu, Goldreich, \& Stevenson 2008),
and it is not known whether the convective interior supports
such banded streams.  Thus, whereas the solar problem is
tightly confined, and the deep convective envelope
with a monotonic surface rotation profile is at least 
a well-motivated problem, this particular
application is truly ``untethered.''
These are purely formal solutions to the TWE.

We consider a surface profile of the form
\beq\label{bandit}
\Omega_0 = {\rm Const.} + \sin(4\pi\sin^2\theta_0).
\eeq
In figures 9 and 10, we show a suite of four representative results 
with the equation (\ref{sin2}) parameters,
$\beta_1 = 0.2, 0.4, 0.6, 0.8$ and $\beta_2 = 0$.
As $\beta_1$ increases, the departure of the isorotation 
contours from Taylor-Proudman columns is evident, until
by $\beta_1=0.8$ many contours show relatively little variation
in $y$.
Closer to the equator, however, the contours retain their 
cylindrical form.  As noted in BBLW, rapid rotators
are probably characterized by relatively small $\beta_1$, because of the 
$1/\Omega^2$ scaling of $f'(\Omega^2)$.  

These results can be compared with numerical simulations of convectively
driven differential rotation in Jupiter, Saturn, Uranus and Neptune.
Such models have grown progressively more realistic, moving from mean-field
(Rieutord et al.\  1984) to Boussinesq (e.g Aurnou \& Heimpel
2004; Aurnou, Heimpel \& Wicht 2007), and most recently to anelastic (Jones \&
Kuzanyan 2009) approximations.  In the latter calculations, 
the Taylor-Proudman constraint is dominant,
and the zonal velocity tends accordingly to be constant on cylinders.

\begin{figure}
\epsfig{file=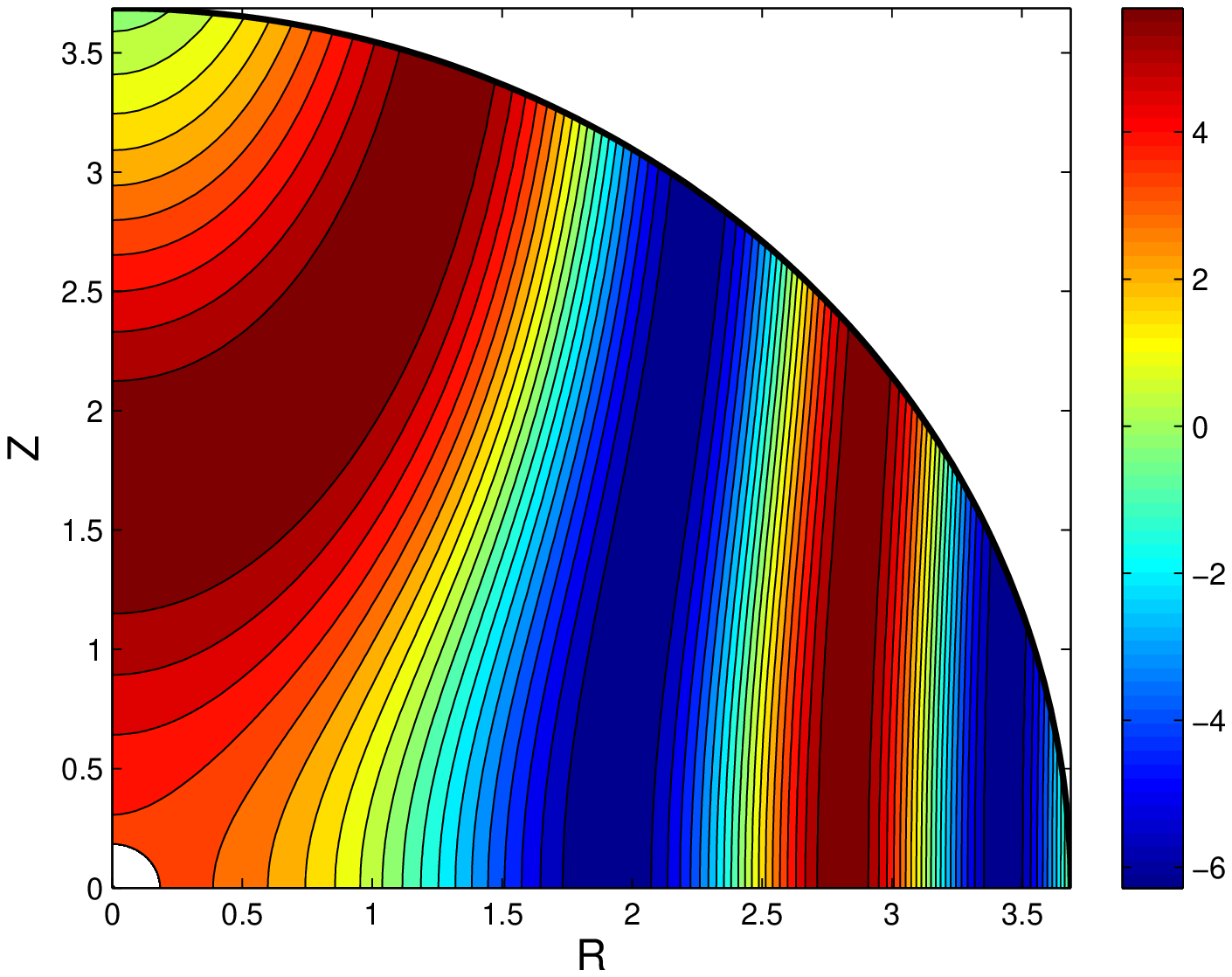, width=8 cm}%,angle=-90}
\epsfig{file=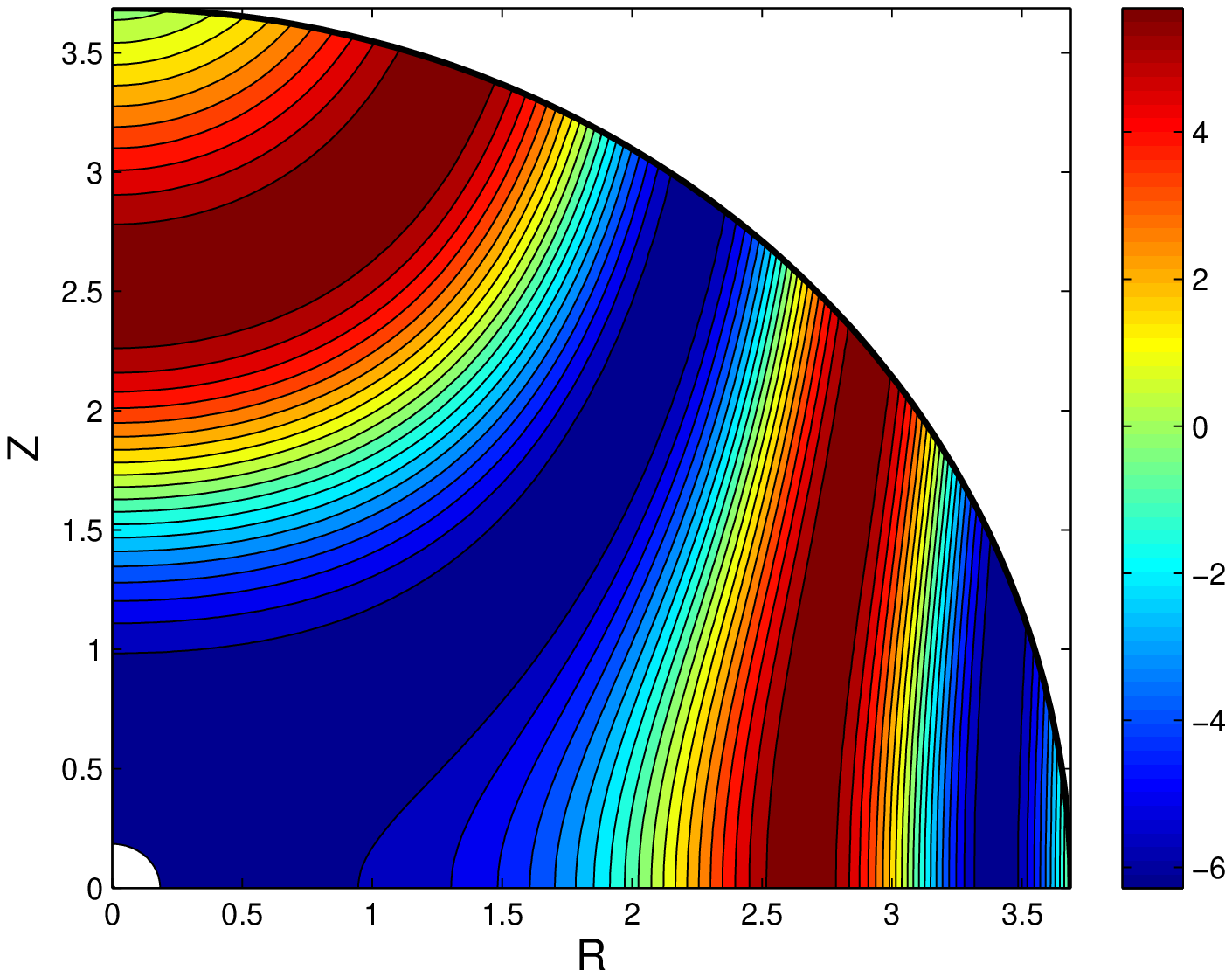, width=8 cm}%,angle=-90}
\caption{Banded surface profile given by equation
(\ref{bandit}). The abscissa $R$ and
ordinate $Z$ are given
in units of the Lane-Emden radius $a$ (eq. [\ref{lea}]).
$\beta_1=0.2$, $\beta_2=0$ (left); 
$\beta_1=0.4$, $\beta_2=0.0$ (right).} 
\end{figure}
\begin{figure}
\epsfig{file=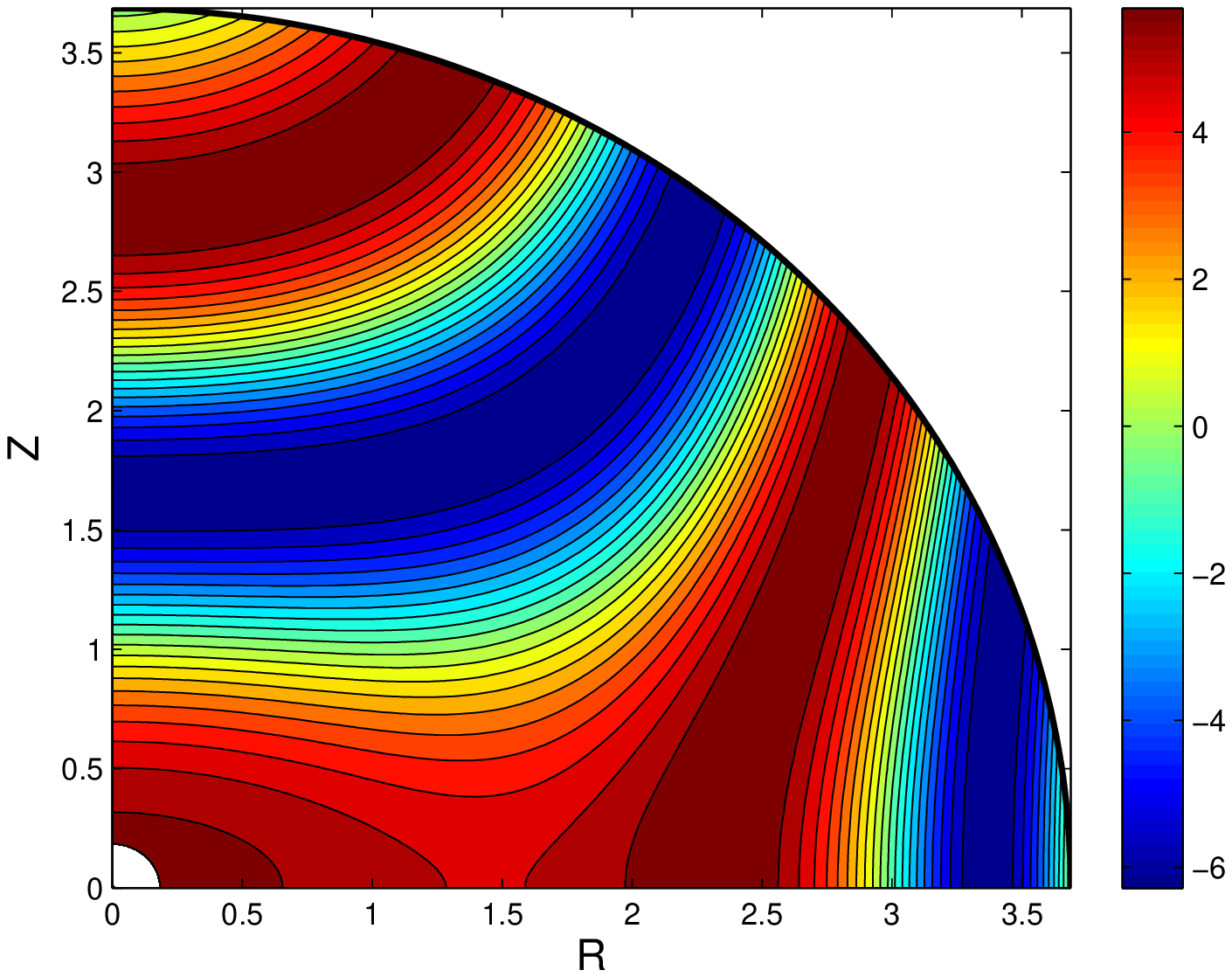, width=8 cm}%,angle=-90}
\epsfig{file=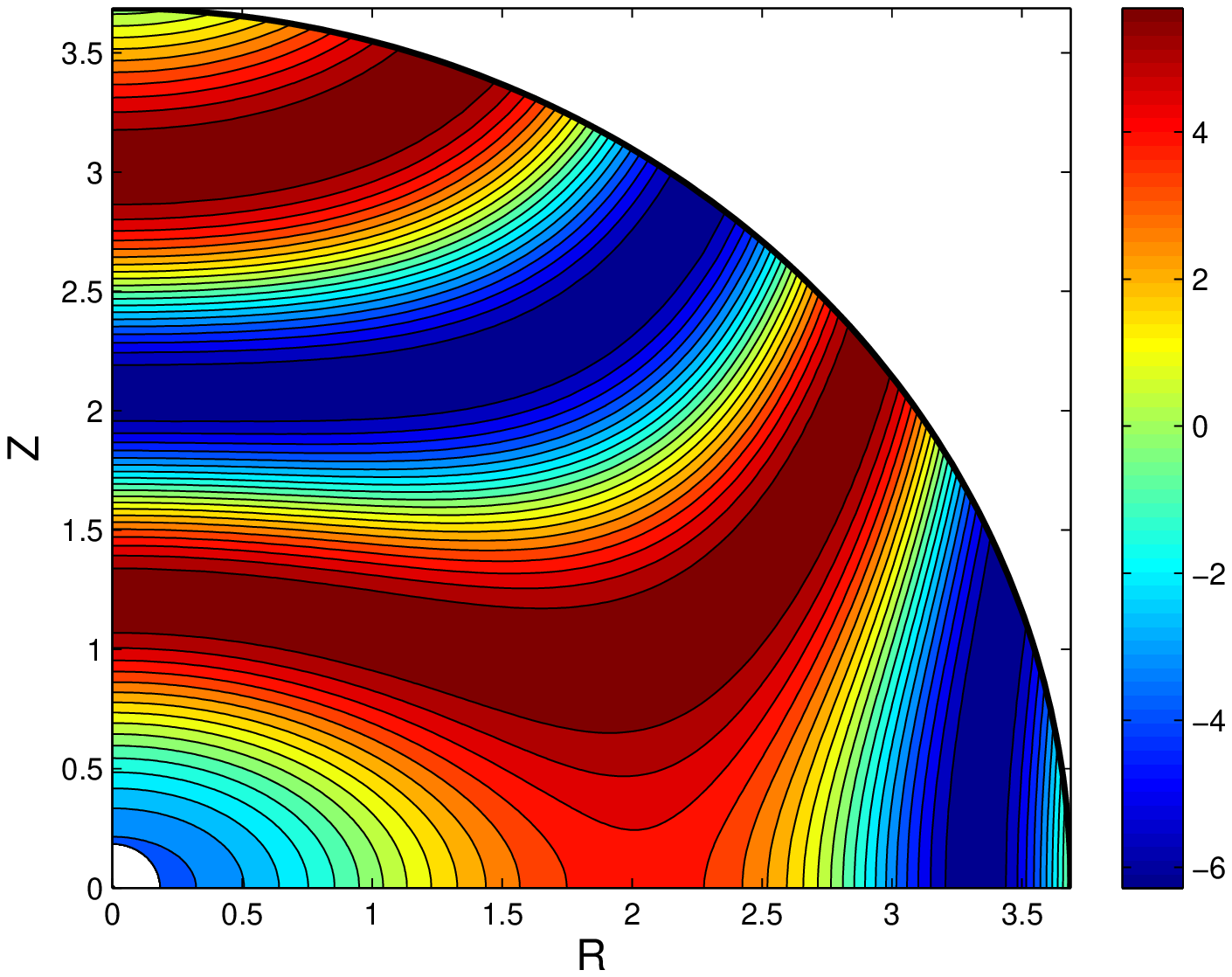, width=8 cm}%,angle=-90}
\caption{As in figure (9), with 
$\beta_1=0.6$, $\beta_2=0.0$ (left); 
$\beta_1=0.8$, $\beta_2=0.0$ (right).} 
\end{figure}

%Functional relationship is maintained by 1 D symmetry ,
%vigorous convection in the quasiradial zone.
\section {Discussion}

Our principal result, embodied in equation (\ref{result}),
is one of remarkable simplicity.  If, in a convecting
star, thermal wind balance holds and
entropy is well mixed in surfaces of constant angular velocity,
the form of the constant $\Omega$ surface passing through a point in the
$r$, $\theta$ plane depends only upon the gravitational potential
and a single ``adjustable'' constant.  This constant is, in principle,
determined from fundamental dynamics, but in practice it depends upon 
knowing how to treat the details of convective turbulent transport.  
One of the important consequences of this work, however,
is precisely this point:  ignorance of the details of turbulent
transport does {\em not} mean complete ignorance of the 
consequences of the transport---in this case, the
shapes of the isorotation surfaces. 

Unfortunately, unlike helioseismology, the fields
of astero- and planetary 
seismology are still very much in their infancies.  
There are no data of quality comparable to the 
solar results with which we
may compare our computed profiles.  
Until such data are available,
the practical utility of our results will primarily 
be as an aid for understanding numerical simulations, and for
theoretical investigations of convective processes 
in which a average equilibrium rotation
needs to be specified or understood (e.g. stellar evolution
and dynamo theories).  

In principle, large scale numerical computations of fully convective stars
are amenable to development at a level evinced by the solar convection studies
(see, e.g. Miesch \& Toomre 2009).
But in practice, this class of calculation is 
extremely demanding of resources, 
so that only the SCZ has been explored in detail and under a large
variety of physical assumptions.   
Recent work, however, has shown that
numerical simulations of convective envelopes are viable, and 
properly interpreted, a potentially rich source of dynamical information
(e.g. Browning 2008; Brun \& Palacios 2009 and references therein).
The investigation of Brun \& Palacios (2009) centered on four models
of red giant stars.  The more slowly rotating envelopes produced
quasi-spherical shells for the isorotational surfaces,
while the more rapidly rotating models showed
classic Taylor-Proudman cylinders.  Our calculations also show
this trend; indeed, strikingly so.  More troublesome,
for a direct
application of the work presented in the current paper,
is the fact that our assumption of thermal wind balance
was not a particularly accurate approximation to the vorticity equation
in the simulations:
both turbulent fluctuations and meridional circulation appeared
to be more important in the red giant calculations than in the 
numerical simulations of the SCZ.  Neither is convective
Rossby number small for the simulations, as it is for the Sun.
Yet the gross features
of the red giant simulations seem rather well captured by our
approach, and this is not because the models lack predictive 
power: inapproprately
changing the sign of $F$, for example, destroys even the 
qualitative agreement with simulations in both the solar
and red giant studies. 
Thus the marginality of our assumptions in the case
of the latter may be more of
a technical concern than a real conceptual impediment for
understanding the qualitative structure of the flow.  But
clearly much more can be done.

Another issue that we have not touched on
is the effect of magnetic fields, shown in an extreme form by Browning's
(2008) model of a fully convective M-star, where magnetic stresses
transform a cylindrical $\Omega$-profile into almost uniform rotation.
There is in fact an interesting analogy between the roles of residual
entropy and of magnetic fields in coupling differentially rotating
cylinders (Taylor 1963; Fearn 2007).

Unlike the compelling fit of the solar isorotational contours
presented in BBLW, the results presented in the current work 
must wait before observational comparisons are possible.
Comparison with simulations is encouraging, however, and shows
promise of genuine progress on both the analytic
and numerical fronts.   In it is particularly noteworthy, for
example, that the poleward-increasing rotation profiles associated
with at least some classes of fully convective stars show
concave internal isorotation contours in simulations.  
We are in the early days yet of a field that is both rapidly
and fruitfully developing.
Even if only a small class of fully convective systems has isorotation
contours described by equation (\ref{result}), this would be
extremely useful.  Not only would a piece of stellar hydrodynamics
be secured, but a point of departure for more complex conditions
would be established.  

%The set of
%``solar approximations''
%that are important to our solutions---a low convective Rossby
%number coupled with a dominant thermal wind balance---are not
%unduly restrictive.  A natural venue for these solutions would
%be compact convective regions such as M dwarfs or the cores
%of massive stars, since these are more likely to be rapid rotators.  

\section*{Acknowledgements}
It is a pleasure to thank M. Browning, 
E. Dormy, H. Latter, and M. Miesch for useful conversations.  
An anonymous
referee provided many useful suggestions that greatly
improved the presentation.  SAB is grateful to the
Max-Planck-Institut f\"ur Astronomie and to the
Dynamics of Discs and Planets programme of the
Isaac Newton Institute for Mathematical Sciences for their
support at the early stages of this work.  
This work has been supported by a grant from the 
Conseil R\'egional de l'Ile de France.

\section*{Appendix I: Pad\'e approximants for $\bb{\Theta_{3/2}(\xi)}$.}

For the calculations presented in this paper, 
we have adopted the following Pad\'e approximant for the 
$n=3/2$ Lane-Emden function $\Theta_{3/2}(\xi)$ (Pascual 1977): 
\beq
\Theta_{3/2}(\xi) \simeq { 1-a_1\xi^2 + a_2\xi^4\over 1+b_1\xi^2+b_2\xi^4},
\eeq
where
\beq
a_1 =  8.46561\times 10^{-2}, \qquad a_2= 8.156966\times 10^{-4},
\eeq
and
\beq
b_1 = 8.201058\times 10^{-2},\qquad b_2 = 1.984127\times 10^{-3}.
\eeq
A simpler, and only slightly less accurate approximation is given by
\beq
\Theta_{3/2}(\xi)\simeq {1-a\xi^2\over 1 +b\xi^2},
\eeq
where 
\beq
a=  7.490699\times 10^{-2}, \qquad b=9.175968\times 10^{-2}.
\eeq
The $a$ and $b$ constants are chosen so that the rational function goes to zero
precisely at $\xi=\xi_0=3.65375$, and to $1-\xi^2/6$ for small $\xi$.  
Contour plots produced by this approximation are almost indistinguishable
from those shown in the paper.  An advantage of this form for $\Theta_{3/2}
(\xi)$ is that it is easily invertible:
\beq
\xi^2 = {1-\Theta\over a + b\Theta} \equiv \Lambda (\Theta),
\eeq
a desirable quality in some applications.  (We have suppressed the subscript
$3/2$ on $\Theta$ for clarity of presentation.)   This defines the
$\Lambda$ function used in equation (\ref{z02}).

\section*{Appendix II: Residual entropy bound}

The mathematical identity
\beq
\left(\dd\sigma'\over \dd r\right)_\theta = 
- \left(\dd\theta\over\dd\ln r\right)_{\sigma'}
\left( {1\over r} {\dd\sigma'\over \dd\theta}\right)_r
\eeq
may be used to establish a lower limit to the radial entropy
gradient in the SCZ.  Following the notion
that constant $\sigma'$ and constant $\Omega$ surfaces coincide,
the first partial derivative on the right side of the equation
may be read off directly from the helioseismology data.  The second
partial derivative on the right is evaluated via the TWE.  
The result is
\beq
\left(\dd\sigma'\over \dd \ln r\right)_\theta =
-2\gamma \tan\theta \left(\dd\ln\Omega\over\dd\ln z \right)_R
\left(\dd\theta\over\dd\ln r \right)_{\sigma'}
\left(r\Omega^2\over g\right),
\eeq
where $g$ is the gravitational field of the SCZ, taken here to be
$2.74\times 10^4$ cm$^2$ s$^{-1}$.  Using recent GONG data 
(kindly supplied to us by R.~Howe), we estimate the logarithmic
$\Omega$ derivative as $-0.12$ and the $\theta$-$\ln r$ gradient as
$-0.3$.  Then with $r=5.95\times 10^{10}$ cm (or 0.85 solar radii),
$\Omega=2.5\times 10^{-6}$ s$^{-1}$, and $\gamma=1.67$, we obtain
\beq
\left(\dd\sigma'\over \dd \ln r\right)_\theta =
-1.6\times 10^{-6} \tan\theta.
\eeq
The numerical values correspond to regions of the SCZ near $\tan\theta=1$.  
Since $\sigma=\sigma'+\sigma_r$, this number is likely to be a lower
bound (in magnitude) to the true midlatitude solar radial entropy gradient
$\dd\sigma/\dd \ln r$.  
Mixing length theory gives values between $10^{-6}$ and $10^{-5}$ for
the absolute value of this gradient (e.g. Stix 2004).

\end{document}